\documentclass[twocolumn,pre]{revtex4}
\usepackage{graphicx}
\graphicspath{{SI_revisions/}}
\usepackage{amsmath}
\usepackage{amsfonts}
\usepackage{amssymb}
\usepackage{dsfont}
\usepackage{enumerate}
\usepackage{booktabs}
\usepackage{xcolor}

\newcommand{\beq}{\begin{equation}}
\newcommand{\eeq}{\end{equation}}

\newcommand{\threep}{$3^{\,\prime}$\,}
\newcommand{\fivep}{$5^{\,\prime}$\,}

\newcommand{\ZS}[1]{\textcolor{black}{#1}}
\newcommand{\YE}[1]{{\color{black}#1}}
\newcommand{\TM}{\color{black}}

\newcommand{\namealgo}{OLGA}

\begin{document}

\title{\namealgo: fast computation of generation probabilities of B-
  and T-cell receptor amino acid sequences and motifs}
\author{Zachary Sethna$^{1}$, Yuval Elhanati$^{1}$, Curtis G. Callan Jr.$^{1,2}$, Aleksandra M. Walczak$^{2*}$,  Thierry Mora$^{3*}$}
\address{	$^{1}$Joseph Henry Laboratories, Princeton University, Princeton, New Jersey 08544 USA \\
$^{2}$Laboratoire de physique th\'eorique, CNRS, Sorbonne
Universit\'e, and \'Ecole Normale Sup\'erieure (PSL University), 24,
rue Lhomond, 75005 Paris, France\\
$^{3}$ Laboratoire de physique statistique, CNRS, Sorbonne Universit\'e, Universit\'e Paris-Diderot,
          and \'Ecole Normale Sup\'erieure (PSL University), 24, rue Lhomond, 75005 Paris, France\\
$^*$ These authors contributed equally.
}

\begin{abstract}
\textbf{Motivation:} 
High-throughput sequencing of large immune repertoires has enabled the development of methods to predict the probability of generation by V(D)J recombination of T- and B-cell receptors of any specific nucleotide sequence. These generation probabilities are very non-homogeneous, ranging over 20 orders of magnitude in real repertoires. Since the function of a receptor really depends on its protein sequence, it is important to be able to predict this probability of generation at the amino acid level. However, brute-force summation over all the nucleotide sequences with the correct amino acid translation is computationally intractable. The purpose of this paper is to present a solution to this problem.\\
\textbf{Results:} We use dynamic programming to construct an efficient and flexible algorithm, called OLGA (Optimized Likelihood estimate of immunoGlobulin Amino-acid sequences), for calculating the probability of generating a given CDR3 amino acid sequence or motif, with or without V/J restriction, as a result of V(D)J recombination in B or T cells. We apply it to databases of epitope-specific T-cell receptors to evaluate the probability that a typical human subject will possess T cells responsive to specific disease-associated epitopes. The model prediction shows an excellent agreement with published data. We suggest that OLGA may be a useful tool to guide vaccine design.\\
\textbf{Availability:} Source code is available at \url{https://github.com/zsethna/OLGA}\\

\end{abstract}
\maketitle

\section{Introduction}

The ability of the adaptive immune system to recognize foreign peptides, while avoiding self peptides, depends crucially on the specificity of receptor-antigen binding and the diversity of the receptor repertoire. Immune repertoire sequencing (Repseq) of B- and T-cell receptors (BCR and TCR) \citep{Six2013,Woodsworth2013b,Lindau2017,Heather2017} offers an efficient experimental tool to probe the diversity of full repertoires in healthy individuals \citep{Weinstein2009,Robins2009,Freeman:2009fja,Robins2010,Mora2016e, Pogorelyy2017CP,Howie2015}, in cohorts with specific conditions \citep{Vollmers:2013bm, Jiang:2013dm, Horns:2017kl, Emerson2017, Pogorelyy2018eLife, Faham2017, Komech2018, Dewitt2018} and evaluate the response to specific fluorescent MHC-multimers \citep{Dash:2017go, Glanville:2017js}. Recent work has shown that responding clonotypes often form disjoint clusters of similar amino acid sequences, which has lead to the identification of responsive amino acid motifs \citep{Dash:2017go, Glanville:2017js}. In order for these techniques to have practical applications in therapy and vaccine design, one needs a fast and efficient algorithm to evaluate which specific amino acid sequences and sequence motifs are likely to be generated and found in repertoires. We present a solution to this problem in the form of an algorithm and computational tool, called {\namealgo}, which implements an exact computation of the generation probability of any BCR or TCR sequence (nucleotide or amino acid), or motif. 

BCR and TCR are stochastically generated by choosing a germline genetic template in each of several cassettes of alternates (V, (D), or J) and then splicing them together with random nucleotide deletions and insertions at the junctions.  
Given a generative model, one can define the generation probability of any nucleotide sequence as the sum of the probabilities of all the generative events that can produce that sequence \citep{Murugan2012, Elhanati2015,Yuval2016, Marcou2018}.  However, computing  the generation probability of amino acid sequences by summing over all consistent nucleotide sequences is impractical: because of codon degeneracy, the number of nucleotide sequences to be summed grows exponentially with sequence length. {\namealgo} is powered by an efficient dynamic programming method to exactly sum over generative events and obtain net probabilities of amino acid sequences and motifs.

We validate our algorithm by comparing its results and performance to Monte-Carlo sampling estimates.  We present results using publicly available data for both TCR $\alpha$ (TRA, \citet{Pogorelyy2017CP}) and $\beta$ (TRB, \citet{Robins2010}) chains and BCR heavy chains (IGH, \citet{DeWitt2016} of humans), and TRB of mice \citep{Sethna:2017bv}. We applied {\namealgo} to a TCR database that catalogs the different CDR3 amino acid sequences responding to a variety of different epitopes associated with disease \citep{Shugay:2018ir}. We computed the generation probability of particular CDR3 amino acid sequences, as well as the net generation probability of all the TCR that respond to a particular epitope. Finally, we discuss {\namealgo}'s applications in vaccine design and other therapeutic contexts.

\section{Methods}

\subsection{Stochastic model of VDJ recombination}

V(D)J recombination is a stochastic process involving several events (gene template selection, terminal deletions from the templates, random insertions at the junctions), each of which has a set of possible outcomes chosen according to a discrete probability distribution. The probability $P^{\rm rec}_{\rm gen}(E)$ of any generation event $E$, defined as a combination of the above-mentioned processes is, for the TRB locus:
\beq\label{genmodel}
\begin{split}
&P^{\rm rec}_{\rm gen}(E) = P_{\rm V}(V)P_{\rm DJ}(D, J) P_{\rm delV}(d_V | V) P_{\rm delJ}(d_J|J)  \\
&\ \times P_{\rm delD}(d_D, d'_D|D) P_{\rm insVJ}(\ell_{\rm VD})p_0(m_1)\left[\prod_{i=2}^{\ell_{VD}} S_{\rm VD}(m_{i}|m_{i-1})\right]  \\
&\ \times P_{\rm insDJ}(\ell_{\rm DJ})q_0(n_{\ell_{DJ}})\left[\prod_{i=1}^{\ell_{DJ}-1}S_{\rm DJ}(n_i|n_{i+1})\right],
\end{split}
\eeq
where $(V,D,J)$ identify the choices of gene templates, $(d_V,d_D,d'_D,d_J)$ are the numbers of deletions at each end of the segments, and $(m_1,\ldots,m_{\ell_{VD}})$ and $(n_1,\ldots,n_{\ell_{DJ}})$ are the untemplated inserted nucleotide sequences at the VD and DJ junctions. These variables specify the recombination event $E$, and are drawn according to the probability distributions ($P_{\rm V}$, $P_{\rm DJ}$, $P_{\rm delV}$, $P_{\rm delD}$, $P_{\rm delJ}$, $P_{\rm insVJ}$, $P_{\rm insDJ}$, $p_0$, $q_0$, $S_{\rm VD}$, $S_{\rm DJ}$). The inserted segments are drawn according to a Markov process starting with the nucleotide distribution $p_0$ and with the transition matrix $R$, and running from the 5' side (left to right) for the VD segment, and from the 3' side (right to left) from the DJ segment. Similar models can be defined for the $\alpha$ chain or for BCR chains. Although here we describe the method for TRB only, it is also implemented for other chains in the software.

Since the same nucleotide sequence can be created by more than one specific recombination event, the generation probability of a nucleotide sequence  is the sum of the probabilities of all possible events that generate the sequence:
$
P^{\rm nt}_{\rm gen}(\boldsymbol \sigma) = \sum_{E\to {\boldsymbol \sigma}} P^{\rm rec}_{\rm gen}(E),
$
where the sum is over all recombination events $E$ that produce the sequence ${\boldsymbol \sigma}=(\sigma_1,\ldots,\sigma_n)$.
The probability of generation of an amino acid sequence, ${\bf a}=(a_1,\ldots,a_L)$ is the sum of the probabilities of all nucleotide sequences that translate into the amino acid sequence:
\beq\label{pgenaa}
P^{\rm aa}_{\rm gen}(a_1,\ldots,a_L)=\sum_{{\boldsymbol \sigma}\sim \bf a} P^{\rm nt}_{\rm gen}(\sigma_1,.,\sigma_{3L})=\!\!\sum_{E\to {\boldsymbol \sigma}\sim {\bf a}}  P^{\rm rec}_{\rm gen}(E),
\eeq
where the $\sim$ sign indicates that ${\boldsymbol \sigma}$ translates into ${\bf a}$. 
We can generalize this approach to any scheme that groups nucleotide triplets, or codons, into arbitrary classes, which we still denote by ${\boldsymbol \sigma}\sim{\bf a}$. In the formulation above, these classes simply group together codons with the same translation according to the standard genetic code.
In an example of generalization, all codons that code for amino acids with a common chemical property, e.g. hydrophobicity or charge, could be grouped into a single class. In that formulation, $(a_1,\ldots,a_L)$ would correspond to a sequence of symbols denoting that property. More generally, any grouping of amino acids can be chosen (including one where any amino acid is acceptable), and the partition can be position dependent. Thus, the generation probability of arbitrary ``motifs'' can be queried. In the following, for ease of exposition, we restrict our attention to the case where ${\bf a}$ is an amino acid sequence.

\subsection{Dynamic programming computation of the generation probability of amino acid sequences}
We now \YE{give an overview of} how {\namealgo} computes Eq.~\ref{pgenaa} without performing the sum explicitly, using dynamic programming. \YE{Fig.~S1-S2 give a graphical overview of the method, and details of the method implementation can be found in SI Secs. I and II and in the code manual.}
Given the genomic nucleotide sequences of the possible gene templates, together with a specific model of the type described in Eq.\,\ref{genmodel}, the algorithm computes the net probability of generating a recombined gene with a given CDR3 amino acid sequence under a given set of V and J gene choices. 

\begin{figure}
\begin{center}
\includegraphics[width=.7\linewidth]{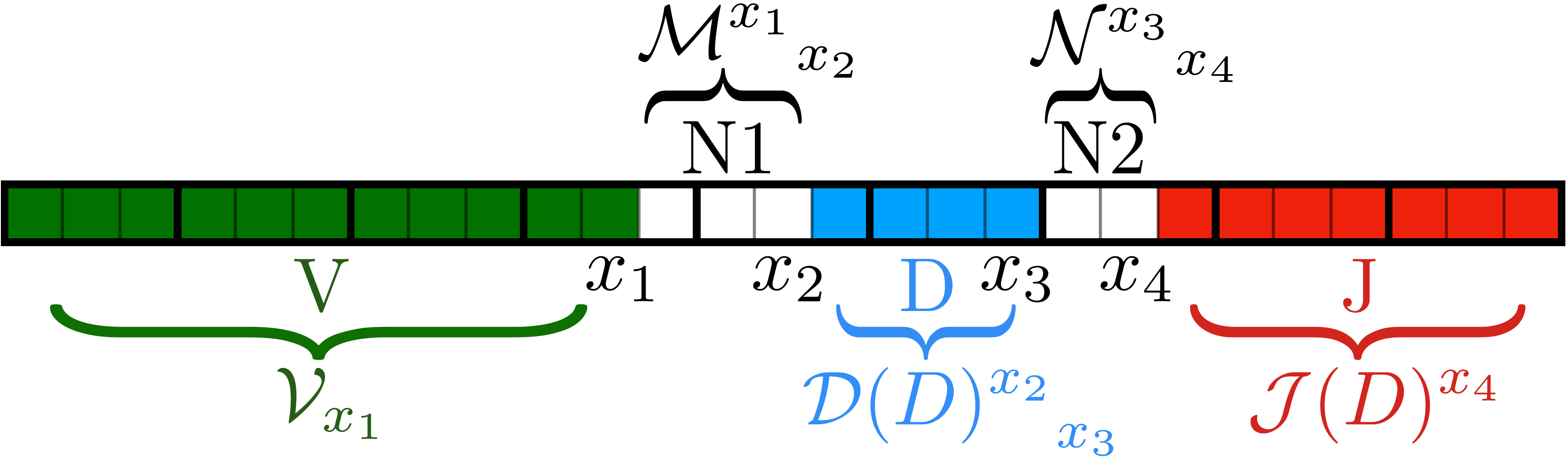}
\end{center}
\caption{Partitioning a CDR3 sequence: boxes correspond to nucleotides and are indexed by integers. Each group of three boxes (identified by heavier boundary lines) corresponds to an amino acid. The nucleotide positions $x_1, \ldots, x_4$ identify the boundaries between different elements of the partition. The $\mathcal{V}$, $\mathcal{M}$, $\mathcal{D}(D)$, $\mathcal{N}$ and $\mathcal{J}(D)$ matrices define cumulated weights corresponding to each of the 5 elements.
}
\label{seq_partition}
\end{figure}

Each recombination event implies an annotation of the CDR3 sequence, assigning a different origin to each nucleotide (V, N1, D, N2, or J, where N1 and N2 are the VD and DJ insertion segments, respectively) that parses the sequence into  5 contiguous segments (see schematic in Fig.\,\ref{seq_partition}).
The principle of the method is to sum over the probabilities of all choices of nucleotides consistent with the known amino acid sequence, over the
possible locations of the 4 boundaries ($x_1$, $x_2$, $x_3$, and $x_4$) between the 5 segments, and over the possible V, D, and J genomic templates (Fig.~\ref{seq_partition}). We do this in a recursive way using matrix operations by defining weights that accumulate the probabilities of events from the left of a position $x$ (i.e. up to $x$), and weights that accumulate events from the right of $x$ (i.e. from $x+1$ on). Specifically, we define the following index notation: $\mathcal{X}_x$ with a subscript called left index, accumulates weights from the left of $x$; $\mathcal{Y}^x$, with a superscript called right index, accumulates weights from the right of $x$; a matrix ${\mathcal{X}^x}_y$ corresponds to accumulated weights from position $x+1$ to $y$ (as will be explained shortly, these objects may have suppressed nucleotide indices as well).
$P^{\rm aa}_{\rm gen}$ is calculated recursively by matrix-like multiplications as:
\begin{equation}\label{pgenmatrix}
P^{\rm aa}_{\rm gen}({\bf a})  =\!\!\sum_{x_1,x_2,x_3,x_4}\!\!\mathcal{V}_{x_1}{\mathcal{M}^{x_1}}_{x_2}  \sum_{D}\left[{{\mathcal{D}(D)}^{x_2}}_{x_3}{\mathcal{N} ^{x_3}}_{x_4}{\mathcal{J}(D)}^{x_4}\right].
\end{equation}
The vector $\mathcal{V}_x$ corresponds to a cumulated probability of the V segment finishing at position $x$; ${\mathcal{M}^x}_y$ is the probability of the VD insertion extending from $x+1$ to $y$; ${\mathcal{N}^x}_y$ is the same for DJ insertions; ${{\mathcal{D}}^x}_y(D)$ corresponds to weights of the D segment extending from $x+1$ to $y$, conditioned on the D germline choice being $D$; ${\mathcal{J}}^x(D)$ gives the weight of J segments starting at position $x+1$ conditioned on the D germline being $D$. This $D$ dependency is necessary to account for the dependence between the D and J germline segment choices \citep{Murugan2012}. All the defined vectors and matrices depend implicitly on the amino acid sequence $(a_1,\ldots,a_L)$, but we leave this dependency implicit to avoid making the notation too cumbersome.

Because we are dealing with amino acid sequences encoded by triplet nucleotide codons, we need to keep track of the identity of the nucleotide at the beginning or the end of a codon. Depending on the position of the index $x$ in the codon, the objects defined above may be vectors of size 4 (or $4\times 4$ matrices) in the suppressed nucleotide index. We use conventions that depend on whether we are considering left or right indices, as follows.

If $x$ is a multiple of 3, i.e. $x=0\ ({\rm mod}\ 3)$, then we do not keep nucleotide information and both $\mathcal{X}_x$ and $\mathcal{Y}^x$ are scalars (whether $x$ is a left or a right index). If $x=1\ ({\rm mod}\ 3)$, then $\mathcal{X}_x$ must be interpreted as a row vector of 4 numbers, $\mathcal{X}_x(\sigma)$, $\sigma=A,T,G,C$, corresponding to the cumulated probability weight that the nucleotide at position $x$ (first position of the codon) takes value $\sigma$. If $x=2\ ({\rm mod}\ 3)$, then $\mathcal{X}_x$ is also a row vector of 4 numbers, $\mathcal{X}_x(\sigma)$, but with a different interpretation: it corresponds to the cumulated probability up to position $x$, with the additional constraint that the nucleotide at position $x+1$ (the last position in the codon) {\em can} take value $\sigma$ (the value is 0 otherwise).
For right indices, the interpretation is reversed and the entries are column vectors: when $x=1\ ({\rm mod}\ 3)$ the $\mathcal{Y}^x$ is a column vector containing the cumulated weights from $x+1$ onwards, with the constraint that the nucleotide at $x$ {\em can} be $\sigma$, and when $x=2\ ({\rm mod}\ 3)$, it is the probability weight that the nucleotide at position $x+1$ {\em is} $\sigma$. Generalizing to matrices, ${\mathcal{X}^x}_y$ is a 4x4, 4x1, 1x4, or 1x1 matrix depending on whether the $x$ and $y$ positions are multiples of 3 or not, with the same rules as for vectors for each type of index.

Entries with left indices are interpreted as row vectors, and entries with right indices as column vectors. Thus, in Eq.~\ref{pgenmatrix} contractions between left and right indices correspond to dot products over the 4 nucleotides when the index is not a multiple of 3, and simply a product of scalars when it is.

The entries of the matrices corresponding to the germline segments, $\mathcal{V}$, $\mathcal{D}(D)$, and $\mathcal{J}(D)$, can be calculated by simply summing over the probabilities of different germline nucleotide segments compatible with the amino acid sequence $(a_1,\ldots,a_L)$ with conditions on deletions to achieve the required segment length. For instance, the $\mathcal{V}$ matrix elements are given by:
\begin{equation}
\mathcal{V}_x(\sigma)=\sum_V P_{\rm V} (V)P_{\rm delV}(l_V\!-\!x) \mathbb{I}(s_x^V\!=\!\sigma) \mathbb{I}(\mathbf{s}^V_{1: x}\sim \mathbf{a}_{1: i}) \textrm{ if }u=1\nonumber
\end{equation}
\begin{equation}
\mathcal{V}_x(\sigma)=\sum_V P_{\rm V} (V)P_{\rm delV}(l_V-x)\mathbb{I}((\mathbf{s}^V_{1: x},\sigma)\sim \mathbf{a}_{1: i}) \textrm{\ \ if }u=2,\nonumber
\end{equation}
\begin{equation}
\mathcal{V}_x=\sum_V P_{\rm V} (V) P_{\rm delV}(l_V-x)\mathbb{I}(\mathbf{s}^V_{1: x}\sim \mathbf{a}_{1: i}) \textrm{\ \ if }u=3,  \label{V_definitions}
\end{equation}
where $x=3(i-1)+u$, i.e. $x$ is the $u^{\rm th}$ nucleotide of the $i^{\rm th}$ codon, $\mathbf s^V$ the sequence of the V germline gene, and $\mathbb{I}$ the indicator function. The $\sim$ sign is generalized to incomplete codons so that it returns a true value if there exists a codon completion that agrees with the motif $\mathbf a$.
Detailed formulas for the other segments are derived using the same principles and are given in the SI Appendix. 
The sums in Eq.~\ref{V_definitions} (and equivalent expressions for J) can be restricted to particular germline genes to compute the generation probability of particular VJ-CDR3 combinations.

The entries of the insertion segment N1 are calculated using the following formula:
\beq\label{insertiontransfer}
{\mathcal{M}^x}_y= P_{\rm insVD}(y-x) L^{u}_{a_{i}} T_{a_{i+1}}\ldots T_{a_{j-1}} R^{v}_{a_{j}},
\eeq
with $y=3(j-1)+v$ (and $x=3(i-1)+u$ as in Eq.\,\ref{V_definitions}).
The transfer matrix
\beq
T_{a}(\tau, \sigma) = \sum_{(n_1,n_2, \sigma) \sim a} S_{\rm VD}(\sigma | n_2)S_{\rm VD}(n_2 | n_1) S_{\rm VD}(n_1 | \tau)
\eeq
corresponds to the probability of inserting a codon coding for $a$ and ending with nucleotide $\sigma$, knowing that the previous codon ended with nucleotide $\tau$. $L^u_a$ and $R^v_a$ are vectors or matrices with different definitions depending on the values of $x$ and $y$ modulo 3, corresponding to the probabilities of inserting incomplete codons on the left and right ends of the insertion segment. Eq.~\ref{insertiontransfer} is only valid for $j>i$, but similar formulas describe the case $i=j$. The precise definitions of $L$ and $R$, the $i=j$ case, and the formulas for $\mathcal{N}$ and the N2 insertion segment, which is exactly equivalent, are all given in detail in the SI Appendix.

The matrix product of Eq.~\ref{insertiontransfer} can be calculated recursively, requiring only $4\times 4$ matrix multiplications. Thus,  all ${\mathcal{M}^x}_y$ elements can be calculated in $\mathcal{O}(L^2)$ operations, instead of the exponential time that would be required using brute-force summation over nucleotides in degenerate codons. Finally, since the sums of Eq.~\ref{pgenmatrix} can also be done recursively through $L\times L$ matrix operations, the whole procedure has $\mathcal{O}(L^2)$ computational complexity.

\section{Results}

\subsection{Method validation}

\begin{figure}
\includegraphics[width=\linewidth]{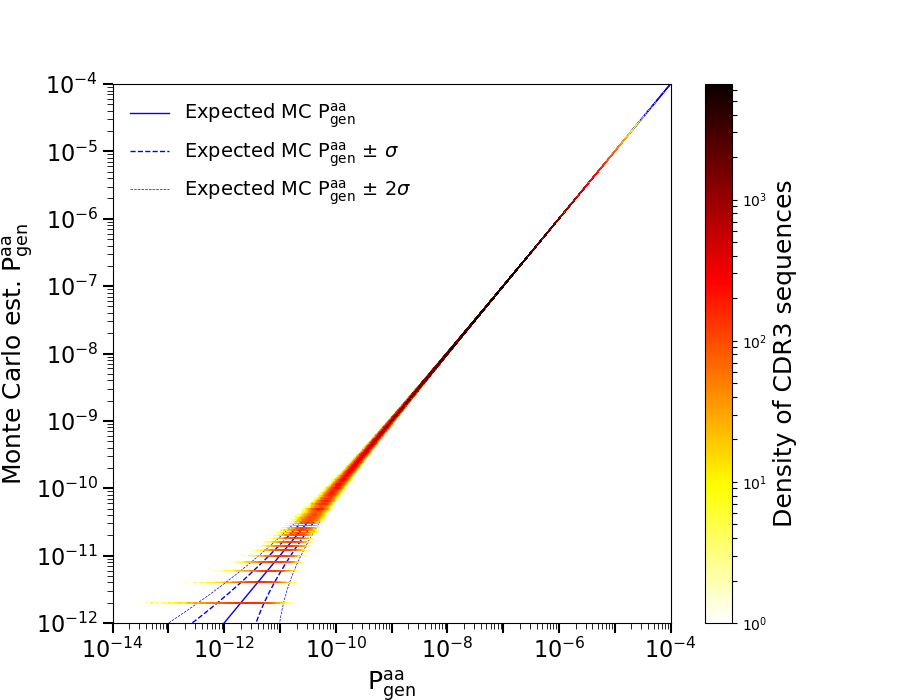}
\caption{Monte Carlo estimate of the generation probability of amino acid CDR3 sequences, $P^{\rm aa}_{\rm gen}$, versus {\namealgo}'s predictions (mouse TRB). The horizontal lines at the lower left of the plot represent CDR3s that were generated once, twice, etc, in the MC sample. The one- and two-sigma curves display the deviations from exact equality between simulated and computed $\rm P_{gen}$ to be expected on the basis of Poisson statistics.}
\label{beta_validation}
\end{figure}

To verify the correctness of the {\namealgo} code, we compared its predictions for generation probabilities to those estimated by Monte Carlo (MC) sequence generation \citep{Pogorelyy2018eLife}. MC estimation is done by drawing events from a given generative model, binning according to the resulting CDR3 amino acid sequence, and normalizing by the total number of recombination events. The scatter plot of the estimated generation probabilities for these sequences against the values predicted by {\namealgo} gives a direct test of the algorithm. As MC estimation is susceptible to Poisson sampling noise, it is important to ensure that enough events are drawn to accurately assess the generative probabilities of individual CDR3 sequences. For this reason, we made the comparison using a generative model inferred from a mouse, rather than human, T cell repertoire, because of the significantly lower entropy of mouse repertoires \citep{Sethna:2017bv}. The specific model was inferred by IGoR  \citep{Marcou2018} using $\sim 70000$ out-of-frame TRB sequences from a mature mouse thymus. MC estimation was done by generating $5 \times 10^{11}$ recombination events, from which the first $10^6$ unique CDR3 amino acid sequences are counted to serve as a sample for the comparison. This procedure provided good sequence coverage, with $>98\%$ of sequences generated at least twice and $>95\%$ of sequences generated at least 10 times. As Fig.\,\ref{beta_validation} shows \YE{for mouse TRB (see Fig. S3 for human TRA)}, MC estimation and {\namealgo} calculation are in agreement (up to Poisson noise in the MC estimate). The Kullback-Leibler divergence between the two distributions, a formal measure of their agreement, is a mere $4.82\times 10^{-7}$ bits. 

\subsection{Comparison of performance with existing methods}
We compared the performance of {\namealgo} to other methods.
Direct calculation of amino acid sequence generation probability using {\namealgo} is orders of magnitude faster than the two possible alternative methods: MC estimation (as described above), or exhaustive enumeration of the generative events giving rise to a given amino acid sequence.
{\namealgo} took 6 CPU hrs to compute the generation probabilities of the $10^6$ amino acid sequences, i.e. 47 seqs/CPU/sec {\TM for mouse TRB (see SI Sec. III and Table S1 for runtimes of other loci)}.
By comparison, MC estimation required 4313 CPU hrs. The scaling for the MC estimation does not depend on the number of queried sequences, but instead is determined by the number of recombinations needed to control the Poisson noise, which scales inversely with generation probability. In practice, to determine the generation probability of a typical sequence (which can be as low $10^{-20}$, see Fig.~\ref{fig:model_pgens} and below), one needs to generate very large datasets, and thus the generation probability of many sequences cannot be calculated by the MC method.

Alternatively, one could list all possible nucleotide sequences that translate to a particular amino acid CDR3 and sum the generation probabilities of each nucleotide sequence, using the IGoR algorithm \citep{Marcou2018}. 
Each amino acid sequence in the mouse validation sample is, on average, coded for by 1.84 billion nucleotide sequences (and much more for human TRB). Since IGoR computes generation probabilities of nucleotide sequences at the rate of $\sim 60$ seqs/CPU/sec, it would take $\sim 8500$ CPU hrs to compute the generation probability of a {\em single} amino acid sequence.  
\YE{A systematic comparison of OLGA with IGoR (Fig.~S4) and MC estimation (Figs.~S4 and S5) as a function of the number of analysed sequences and their CDR3 lengths shows that OLGA is faster than both other methods for all practical purposes (see Sec. IV for details).}

\subsection{Distribution of generation probabilities and diversity}

\begin{figure}
\includegraphics[width=\linewidth]{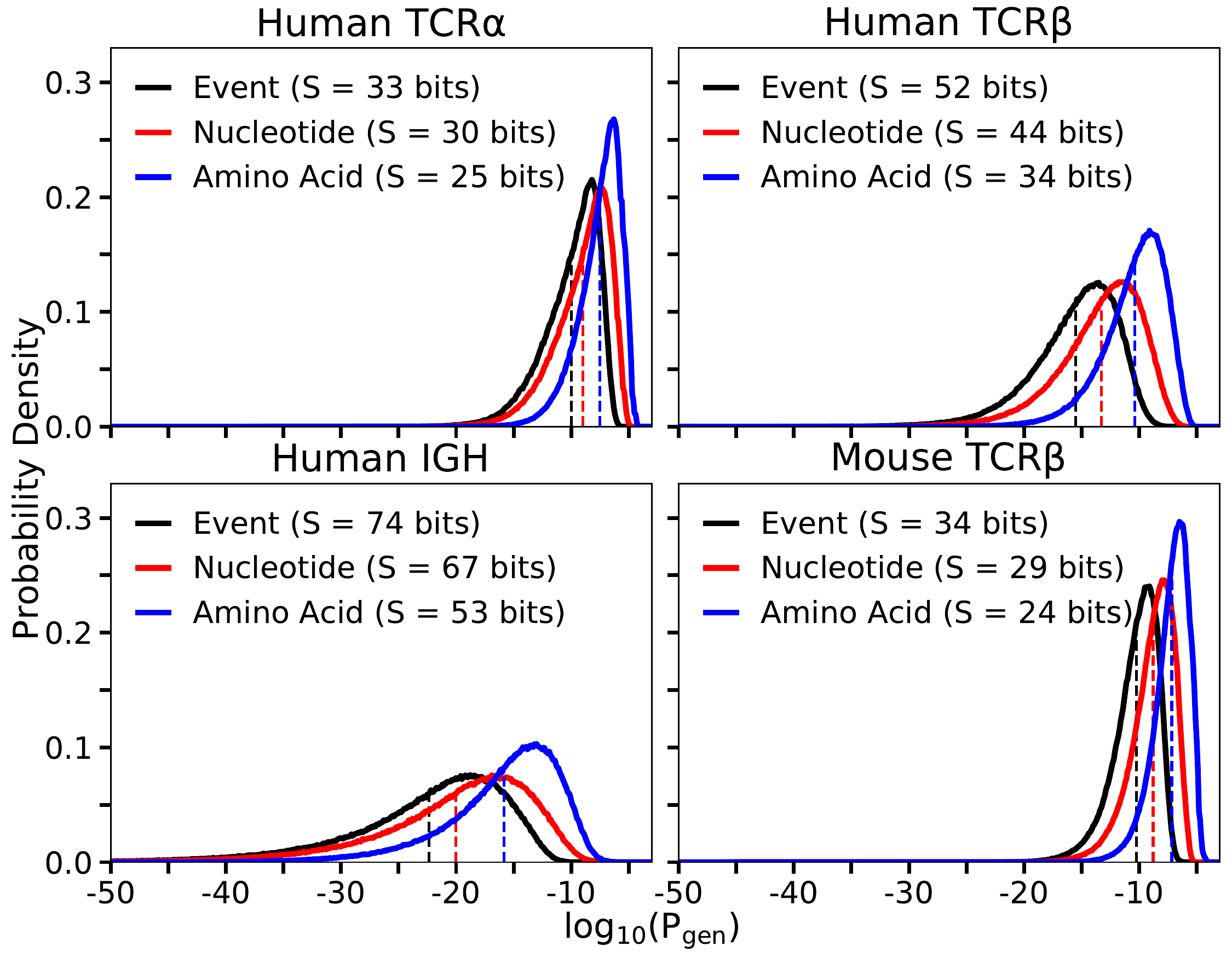}
\caption{Distributions of probabilities of recombination events ($P_{\rm gen}^{\rm rec}$), nucleotide CDR3 sequences  ($P_{\rm gen}^{\rm nt}$), and CDR3 amino acid sequences  ($P_{\rm gen}^{\rm aa}$) in different contexts. Each curve is determined by Monte Carlo sampling of $10^{6}$ productive sequences for the indicated locus, and computing its generation probabilities at the three different levels.
Entropies in bits ($S$) are, up to a $\ln(2)/\ln(10)$ factor, the negative of the mean of each distributions, indicated by dotted lines. 
}
\label{fig:model_pgens}
\end{figure}

V(D)J recombination produces very diverse repertoires of nucleotide sequences, with a very broad distribution of generation probabilities spanning up to 20 orders of magnitude \citep{Murugan2012,Elhanati2015}. This distribution gives a comprehensive picture of the diversity of the process, and can be used to recapitulate many classical diversity measures \citep{Mora2016e}, and to predict the overlap between the repertoires of different individuals \citep{Elhanati:2018gy}. In particular, the opposite of the mean logarithm of the generation probability, $-\langle \log_2 P_{\rm gen}\rangle$, is equal to the entropy of the process. While previous work focused on nucleotide sequence generation, {\namealgo} allows us to compute this distribution for amino acid sequences.

Fig.\,\ref{fig:model_pgens} shows the distribution of $P^{\rm aa}_{\rm gen}$ for 4 loci: human and mouse TRB, human TRA, and human IGH, and compares it to the distributions of nucleotide sequence generation probabilities,  $P^{\rm nt}_{\rm gen}$, and recombination event probabilities, $P^{\rm rec}_{\rm gen}$. \YE{While all these datasets are based on DNA RepSeq, we checked that the generation probability distribution was robust to the choice of protocol by computing the TRB distribution for independent datasets generated by RNA RepSeq \citep{Sims2016,Wu2018,Wang2010} (Figs.~S6 and S7, and SI Sec. V).}
 \YE{The generation models used here and elsewhere in this paper were taken from \citet{Marcou2018}, except for the human TRB model which was relearned using IGoR from one individual in \citet{Emerson2017} as a check.}
Going from recombination events to nucleotide sequences to amino acid sequences leads to substantial shifts in the distribution, and corresponding drops in entropies, as the distribution is progressively coarse-grained.
 \YE{Higher generation probability of a given receptor sequence leads to higher chance of finding it in any given individual. Generation probabilities may be constrasted to the scale set by the inverse of the number of independent recombination events (estimated between $10^8$ \citep{Qi:2014hr} and $10^{10}$ \citep{Lythe:2016fy} for human TCR). Generation probabilities above this limit ($10^{-10}$ to $10^{-8}$ for human TCR) can be considered ``large'' as the corresponding receptor will almost surely exist in each individual \citep{Elhanati:2018gy}. Another relevant scale to distinguish small from large generation probabilities is given by their geometric mean (dashed lines in Fig.~\ref{fig:model_pgens}).
}

{\TM
\subsection{Cross-species generation probabilities}
While distinct species differ in their generation mechanisms, they may yet be able to generate the same CDR3s. Using OLGA, we computed the probabilities of producing human TRB CDR3s by the mouse recombination model, and vice versa (details in SI Sec. VI). An impressive 72.6\% of human CDR3s can theoretically be produced by mice, and 100\% of mouse CDR3s can be produced by humans. While cross-species generation probabilities are lower than intra-species ones (Fig.~S8), they are correlated (Fig.~S9). These results suggest that CDR3s observed in the repertoires of humanized mouse models of human diseases could be relevant for predicting their presence in human repertoires as well. OLGA allows for evaluating this potential, and could be used to inform clinical trials.
}

\subsection{Generation probability of specific TCR}
\begin{figure}
  \includegraphics[width=\linewidth]{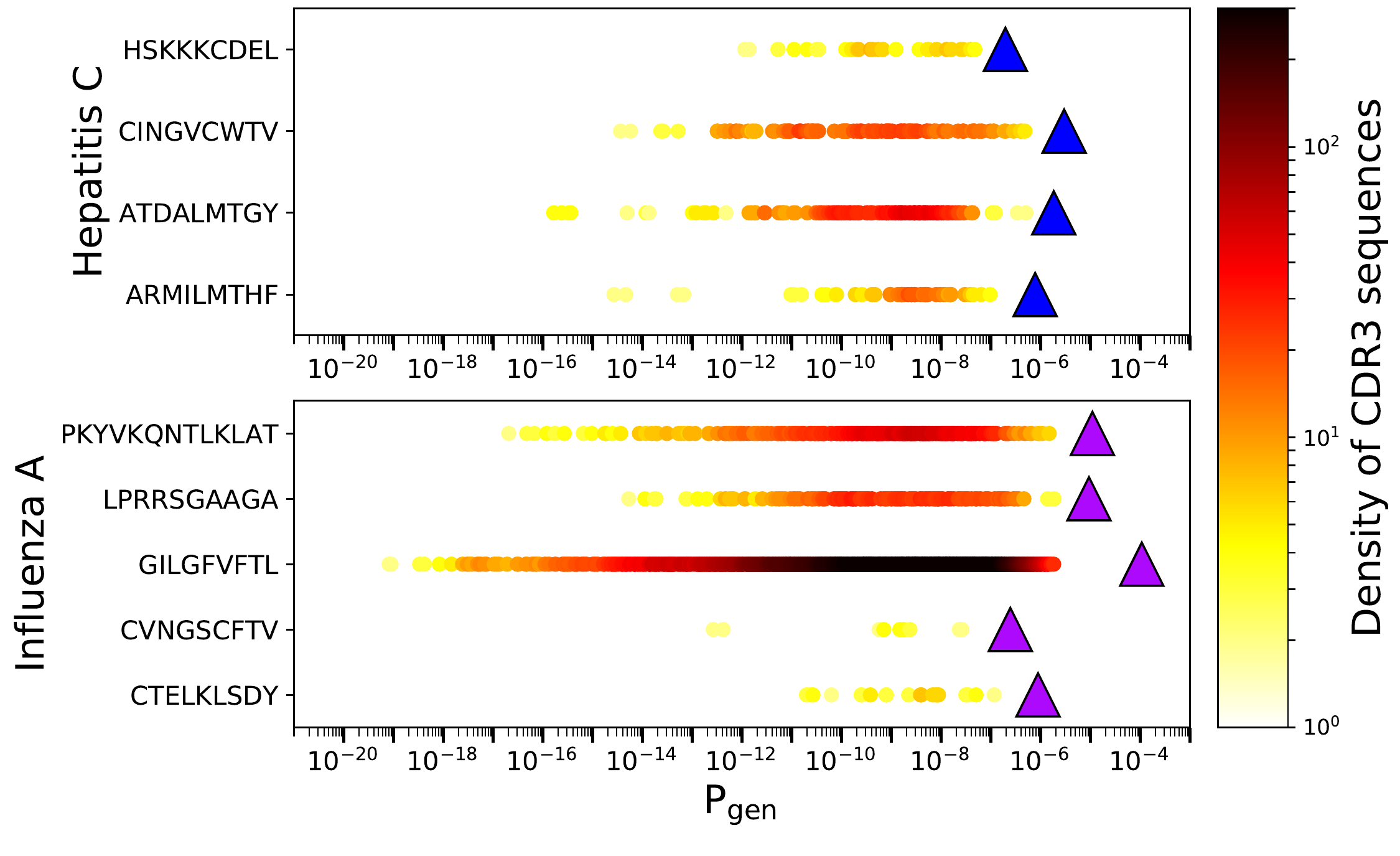} 
  \caption{Generation probabilities of human CDR3s that respond to hepatitis C and influenza A epitopes. $P^{\rm aa}_{\rm gen}$ of sequences that respond to an epitope are plotted as circles (color encodes density of the points). The fraction of the repertoire  specific to each epitope ($P^{\rm func}_{\rm gen}$ as defined in Eq.\,\ref{funPgen} ) is obtained as the sum of the $P^{\rm aa}_{\rm gen}$ for each of the corresponding sequences (values plotted as triangles).}
  \label{fig:epitope_raster}
\end{figure}

We can use {\namealgo} to assess the total fraction of the generated repertoire that is specific to any given epitope, simply by summing the generation probabilities of all \YE{TRB} sequences known to bind specifically to that epitope:
\beq
\label{funPgen}
P^{\rm func}_{\rm gen} ({\rm epitope}) = \sum_{{\bf a}\,|\, {\rm epitope}} P^{\rm aa}_{\rm gen} ({\bf a}),
\eeq
where "${\bf a}\,|\, {\rm epitope}$" means that the amino acid sequence ${\bf a}$ recognizes the epitope. 
Many experiments, based {\em e.g.} on multimer sorting assays \citep{Dash:2017go,Glanville:2017js} or T-cell culture assays, have established lists of epitope-specific TCR sequences for a number of disease-related epitopes. We used the VDJdb database \citep{Shugay:2018ir}, which aggregates such experiments, to compute $P^{\rm func}_{\rm gen}$ of \YE{all TRB known to be reactive against} several epitopes.
In Fig.\,\ref{fig:epitope_raster} we show results for 4 epitopes associated with Hepatitis C, and 5 epitopes associated with Influenza A.
The net fraction of the repertoire specific to these epitopes ($10^{-7}$ to $10^{-4}$) is large in the sense defined above, meaning that any individual is likely to have many copies of reactive T cells in their naive repertoire. 

\begin{figure}
\includegraphics[width=\linewidth]{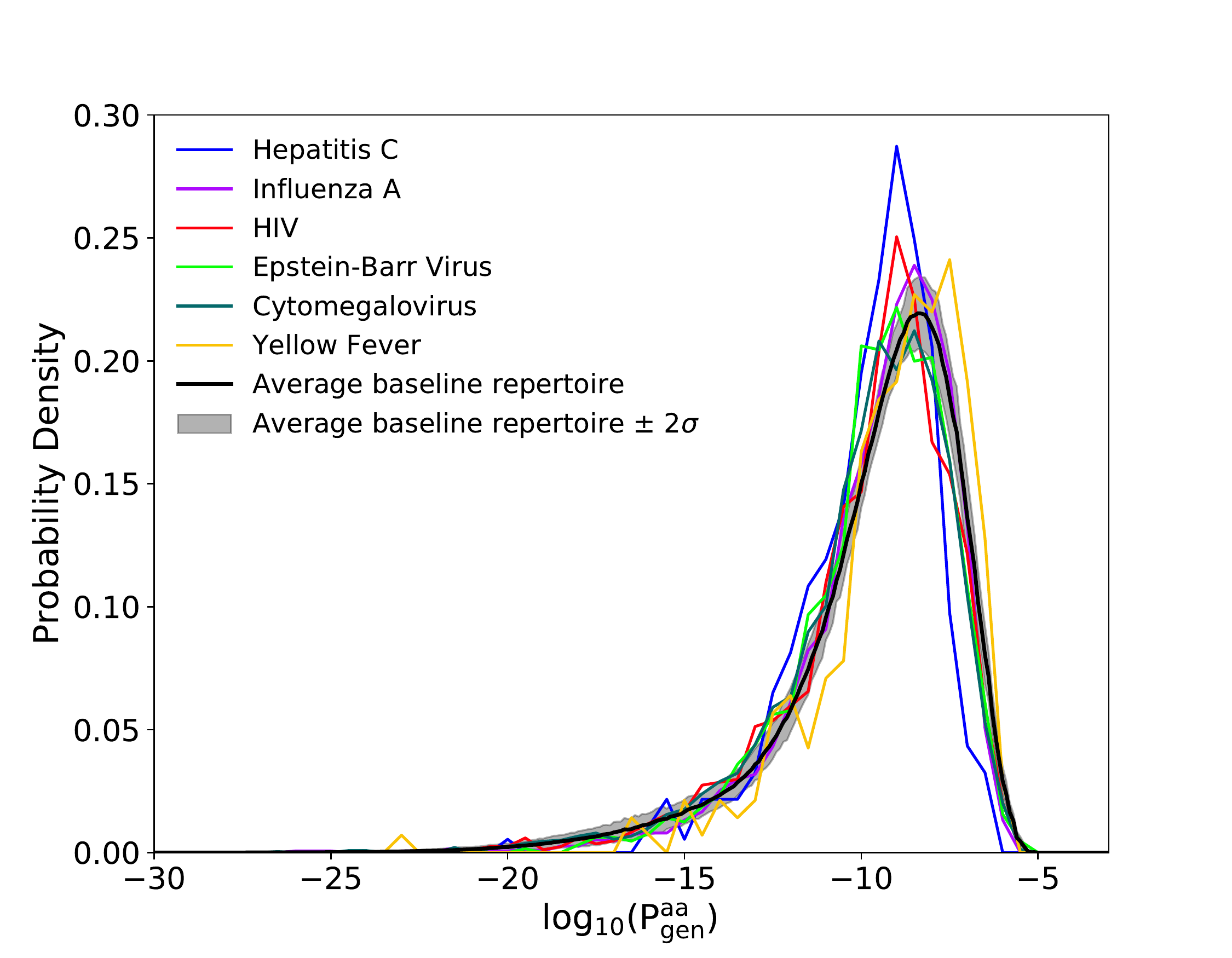}
\caption{Distributions of TRB generation probabilities $P_{\rm gen}^{\rm aa}$ for sequences in the VDJdb database that bind to any epitopes of 6 different viruses (colored curves). For comparison, we plot (black curve) the same distribution for the unsorted TRB repertoire of a typical healthy subject; the $2\sigma$ variance represents biological variability across multiple individuals (data from \citet{Emerson2017}) }
 \label{fig:disease_pgens}
\end{figure}

\YE{The presence of any specific TCR in the repertoire will be affected by the recombination probability of both its $\alpha$ and $\beta$ chains, and also by function-dependent selective pressures. Assessing accurately the fraction of reactive TCRs in the blood is beyond the scope of this method. However, it is still interesting to ask} 
whether epitope-specific \YE{TRB} sequences had higher generation probabilities than regular sequences, either because of observational biases, or because the immune system might have evolved to make them more likely to be produced. To answer that question, we display in Fig.\,\ref{fig:disease_pgens} the $P_{\rm gen}^{\rm aa}$ distribution of the sequences listed in VDJdb that are specific to any epitope of each of 6 commonly studied viruses. For comparison we plot the $P_{\rm gen}^{\rm aa}$ distribution of  the full TRB sequence repertoire of a healthy donor (data taken from \citet{Emerson2017}). 

\YE{The viral distributions are very similar to each other, and also to the healthy repertoire background, meaning that the ability of a CDR3 to respond to a particular disease epitope is not strongly correlated with its generation probability.}
\YE{To see whether this result was confirmed in the case of a real infection, we repeated the same analysis on TRB RepSeq data from T-cells responding to three different types of pathogens (fungus, bacteria, and toxin) \citep{Becattini2015}. Consistently, we found that their distribution of generation probability was identical to that of naive sequences (SI Fig.~S10 and SI Sec. VII).}

\subsection{Model accurately predicts the frequencies of sequences and of groups of specific sequences}
\begin{figure}
\includegraphics[width=\linewidth]{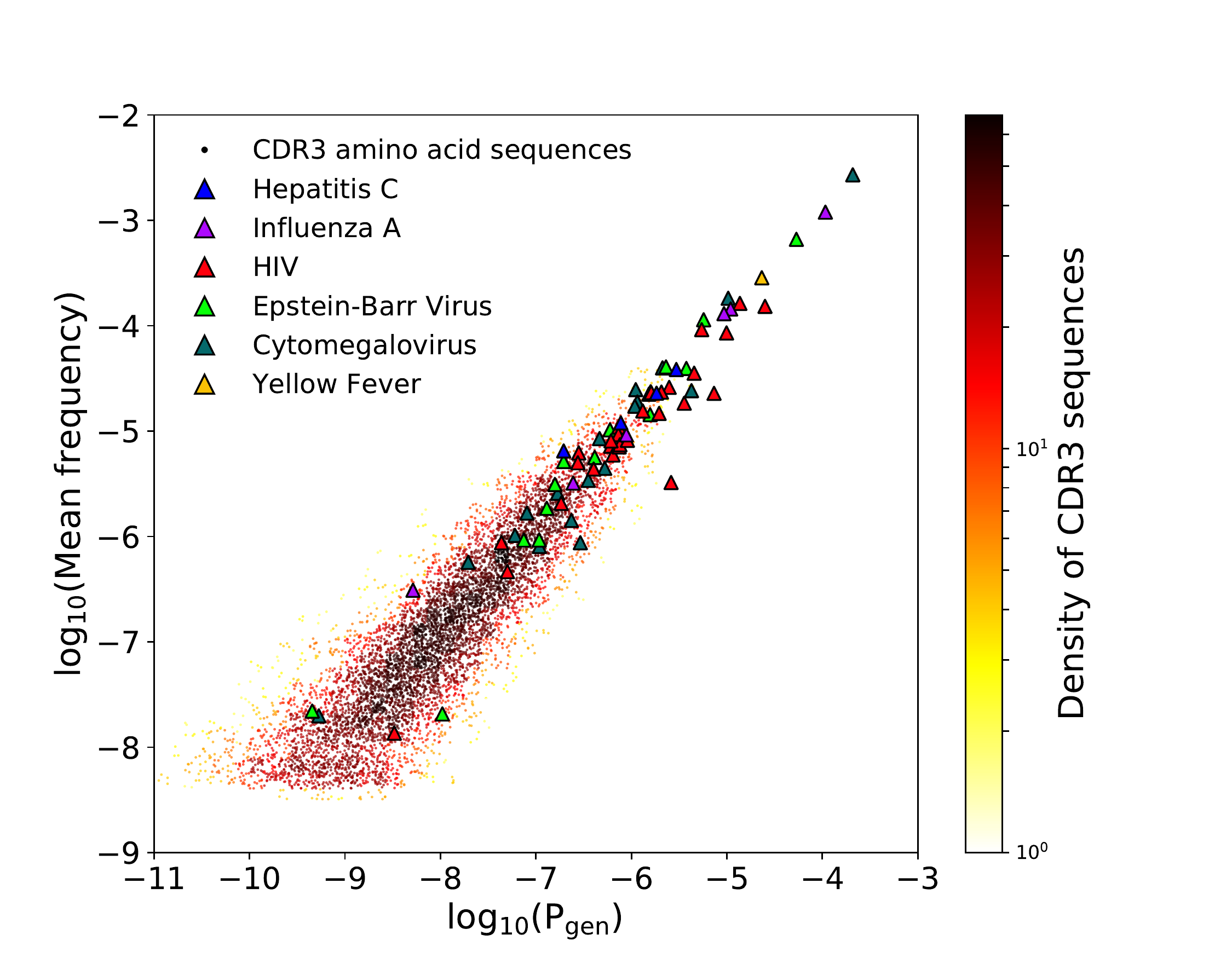} 
\caption{Mean occurrence frequencies across a collection of 658 human samples of all CDR3 sequences in the VDJdb database, plotted against their computed  $P^{\rm aa}_{\rm gen}$ (dots, colored by their density in the plot). Also, the net occurrence frequency in the VDJdb database of epitope-related collections of sequences, plotted against their computed $P^{\rm func}_{\rm gen}$  (triangles, colored to identify the virus the epitope belongs to).}
  \label{fig:epitope_freq_vs_pgen}
\end{figure}

To compare {\namealgo}'s predictions with sequence occurrence frequencies in real data, we used the aggregated TRB repertoire of 658 human subjects described in \citet{Emerson2017} as a test resource. More specifically, we measured the frequencies in this large dataset of the specific CDR3 sequences contained in the VDJdb database \citep{Shugay:2018ir}, and compared them to the values assigned by \namealgo.
When measuring frequencies we discarded read count information, recording only the presence or absence of nucleotide sequences in each individual in order to eliminate effects of clonal expansion and PCR amplification bias, averaging over the 648 individuals in the \citet{Emerson2017} dataset to get reliable estimates of frequencies. Each sequence in the VDJdb database is displayed as a dot  in Fig.\,\ref{fig:epitope_freq_vs_pgen}, and the resulting distribution shows a strong correspondence between mean frequency in the large data set and the predicted $P^{\rm aa}_{\rm gen}$ of that sequence.

We then measured the fraction of CDR3s in the aggregated repertoire that is specific to epitopes associated with 6 viruses (using lists of specific sequences in VDJdb), and compared it to {\namealgo}'s prediction, $P_{\rm gen}^{\rm func}$. The agreement was again excellent (triangles in Fig.\,\ref{fig:epitope_freq_vs_pgen}). 
Again we observe that most epitope-specific sequence groups have large enough frequencies to be found in any individual. Thus, the model can be used to predict the size of repertoire subsets specific to any epitope, as long as specificity data are available for this epitope.

\subsection{Generation probability of sequence motifs}
{\namealgo} can also compute the generation probability of any sequence motif, encoded by a string of multiple choices of amino acids. We apply this feature to calculate the net frequency of epitope-specific motifs, and of motifs that define the TRA sequence of invariant T-cells.

T-cell sequences that can bind a given epitope are often closely related to each other, and this similarity can sometimes be partially captured by sequence motifs. We evaluated the probabilities of motifs derived from a recent study of CDR3 sequence specificity to a variety of epitopes \citep{Dash:2017go}. We took two motifs corresponding to TRA and TRB VJ-CDR3 combinations of TCRs  that are known to bind the Epstein-Barr virus HLA-A*0201-BMLF$_{1280}$ (BMLF) and the influenza virus HLA-A*0201-M$_{158}$ (M1) epitopes. The motifs and generation probabilities are reported in Table\,\ref{tab:table4}. 

\begin{table}
\begin{center}
\caption{Epitope-specific TCR motifs for the Epstein-Barr virus HLA-A*0201-BMLF$_{1280}$ (BMLF) and influenza virus HLA-A*0201-M$_{158}$ (M1) epitopes from \citet{Dash:2017go}, and their generation probabilities. Each motif was associated with specific V/J gene choices. In the motifs we use the conventions: X, any one amino acid; [A..B], any one of the listed amino acids; X\{0,\}, arbitrary amino acid string.}
\label{tab:table4}
\begin{tabular}{|c|c|c|}
\hline
epitope\,:\,chain\,:\,V/J & CDR3 motif  & $P_{\rm gen}$ \\ 
\hline
BMLF\,:\,$\alpha$\,:\,5/31 & CAXD[NSDA]NARLMF  & $1.8 \!\cdot\! 10^{-7}$\\ 
\hline
BMLF\,:\,$\beta$\,:\,20-1/1-2,1-3 & CSARDX[TV]GNX\{0,\}  & $5.1 \!\cdot\! 10^{-7}$ \\
\hline
M1\,:\,$\alpha$\,:\,27/42 & CAXGGSQGNLIF & $2.2 \!\cdot\! 10^{-5}$\\
\hline
M1\,:\,$\beta$\,:\,19/all & CASSXR[S\!A\!][S\!T\!A\!G\!]X[E\!T\!]Q[Y\!F\!]F & $1.7 \!\cdot\! 10^{-6}$\\
\hline
\end{tabular}
\end{center}
\end{table}

As a second application, we estimated the probabilities of generating a TRA chain corresponding to one of the motifs associated with Mucosal associated invariant T cells (MAIT) and invariant natural killer T cells (iNKT). The motifs, which were collected from \citet{Gherardin2016}, and their probabilities are shown in Table \ref{tab:table3}. The relatively high values for these motifs imply that these invariant chains are generated with high frequency in the primary repertoire and shared by all individuals, confirming the conclusions of \citet{Venturi2013}.

\begin{table}[h]
\begin{center}
\caption{Generation probabilities of motifs corresponding to invariant T cell (iNKT and MAIT cells) TRA chain, assembled from serquence in \citet{Gherardin2016}.}
\label{tab:table3}
\begin{tabular}{|c|c|c|c|}
\hline
\rm Type & V/J & \rm CDR3 motif & $\rm P_{gen}$ \\ \hline
\rm iNKT & 10/18 &\rm CVVSDRGSTLGRLYF & $1.26\cdot 10^{-6}$ \\
\rm MAIT & 1-2/33 &\rm CAV[KSM]DSNYQLI[WF] & $1.79\cdot 10^{-5}$ \\
\rm MAIT & 1-2/12 &\rm CAVMDSSYKLIF & $4.71\cdot 10^{-6}$ \\
\rm MAIT & 1-2/20 & \rm CAVSDNDYKLSF & $3.11\cdot 10^{-7}$ \\
\hline
\end{tabular}
\end{center}
\end{table}

\section{Discussion}

Because the composition of the immune repertoire results from a stochastic process, the frequency with which distinct T- and B-cell receptors are generated is a quantity of primary interest. This frequency is computationally difficult to evaluate because each amino acid sequence can be created by a very large number of recombination events. Our tool overcomes that challenge with dynamic programming, allowing it to process $\sim 50$ sequences per second on a single CPU.
In its current state {\namealgo} can compute the probabilities of CDR3 sequences and motifs, with or without V/J restriction, of 4 chain loci (human and mouse TRB, human TRA, and human IGH), but the list can readily be expanded by learning recombination models for other loci and species using IGoR \citep{Marcou2018} which shares the same model format. Obvious additions include the light chains of BCR \citep{Toledano2018}, and more mouse models.
While the algorithm evaluates the probability of single chains, recent analyses show that chain pairing in TCR is close to independent \citep{Grigaityte2017arxiv,Dupic2018arxiv}. The probability of generating a whole TCR receptor can thus be computed by taking the product over the two chains.

{\namealgo} can be used to compute baseline receptor frequencies and to identify outlying sequences in repertoire sequencing datasets. In \citet{Elhanati:2018gy} we used it to shed light on the question of public repertoires ---\,composed of sequences shared by many individuals\,--- and predict quantitatively its origin by convergent recombination \citep{Venturi2008a,Madi2014a,Madi2017}.
Deviations from the baseline expectancy have been used to identify disease-associated TCR from cohorts of patients \cite{Emerson2017,Seay2016,Fuchs2017,Faham2017,Zhao2016}, and to identify clusters of reactive TCRs from tetramer experiments \citep{Glanville:2017js} and vaccination studies  \citep{Pogorelyy2018arxiv}. Such estimates could be made faster and more reliable by {\namealgo}, especially for rare sequences, and without the need for a negative control cohort \citep{Pogorelyy2018eLife}. \YE{In the future, {\namealgo} could be useful in vaccine and therapy design by focusing attention on clonotypes that are likely to be present in every individual.}

We applied {\namealgo} to an experimental database of TCR responding to a variety of disease-associated epitopes. These selected TCR do not differ in their generation probabilities from those of random TCR found in the blood of healthy donors. 
However, some viral epitopes bind a much larger fraction of the repertoire than others. This observation has potentially important consequences for vaccine
design. Since vaccine epitopes stimulate TCR in a pre-existing
repertoire, epitopes targeting receptor sequences that are more
likely to be generated will have a higher success rate in a wider
range of individuals. {\namealgo} can be used to identify such epitopes by computing their specific repertoire fractions, $P_{\rm gen}^{\rm func}$. While our examples are restricted to TCR, {\namealgo} can also handle BCR and could be used to compute the generation probabilities of BCR precursors of highly reactive or broadly neutralizing antibodies, and thus guide vaccine design in that case as well. The algorithm does not yet handle hypermutations, and extending it to include them would be a useful development.

\medskip
{\bf Acknowledgements.} The work of TM and AMW was supported in part by grant ERCCOG n. 724208. The work of ZS and CC was supported in
part by NSF grant PHY-1607612. The work of CC was also supported in part by NSF grant PHY-1734030. The work of YE was supported by a fellowship from the V Foundation. The authors declare no conflicts of interest.

\medskip

\bibliographystyle{natbib}

\onecolumngrid

\clearpage

\appendix

\setcounter{table}{0}
\renewcommand{\thetable}{S\arabic{table}}%
\setcounter{figure}{0}
\renewcommand{\thefigure}{S\arabic{figure}}%

\section{Additional matrix definitions for VDJ algorithm}


Recall that the generative VDJ model is defined as:
\beq\label{genmodel}
\begin{split}
P^{\rm rec}_{\rm gen}(E) = P_{\rm V}(V)P_{\rm DJ}(D, J) P_{\rm delV}(d_V | V) P_{\rm delJ}(d_J|J) P_{\rm delD}(d_D, d'_D|D)&P_{\rm insVJ}(\ell_{\rm VD})p_0(m_1)\left[\prod_{i=2}^{\ell_{VD}} S_{\rm VD}(m_{i}|m_{i-1})\right]  \\
\times &P_{\rm insDJ}(\ell_{\rm DJ})q_0(n_{\ell_{DJ}})\left[\prod_{i=1}^{\ell_{DJ}-1}S_{\rm DJ}(n_i|n_{i+1})\right],
\end{split}
\eeq
with
\beq\label{pgenaa_sum_over_events}
P^{\rm{aa}}_{\rm gen}(a_1,\ldots,a_L)=\sum_{{\boldsymbol \sigma}\sim \boldsymbol a} P^{\rm nt}_{\rm gen}(\sigma_1,\ldots,\sigma_{3L})=\sum_{E\to {\boldsymbol \sigma}\sim {\bf a}} P^{\rm rec}_{\rm gen}(E).
\eeq

As described in the main text, the dynamic programming algorithm can be summarized by the summation over the positions $x_1,\ x_2,\ x_3,$ and $x_4$ of the following matrix multiplication:
\beq\label{pgenaa_dyn_prog_sum}
P^{\rm aa}_{\rm gen}(a_1,\ldots,a_L)  =\sum_{x_1,x_2,x_3,x_4}\mathcal{V}_{x_1}{\mathcal{M}^{x_1}}_{x_2}  \times \sum_{D}\left[{{\mathcal{D}(D)}^{x_2}}_{x_3}{\mathcal{N} ^{x_3}}_{x_4}{\mathcal{J}(D)}^{x_4}\right].
\eeq

\begin{figure}
\begin{center}
\includegraphics[width=.5\linewidth]{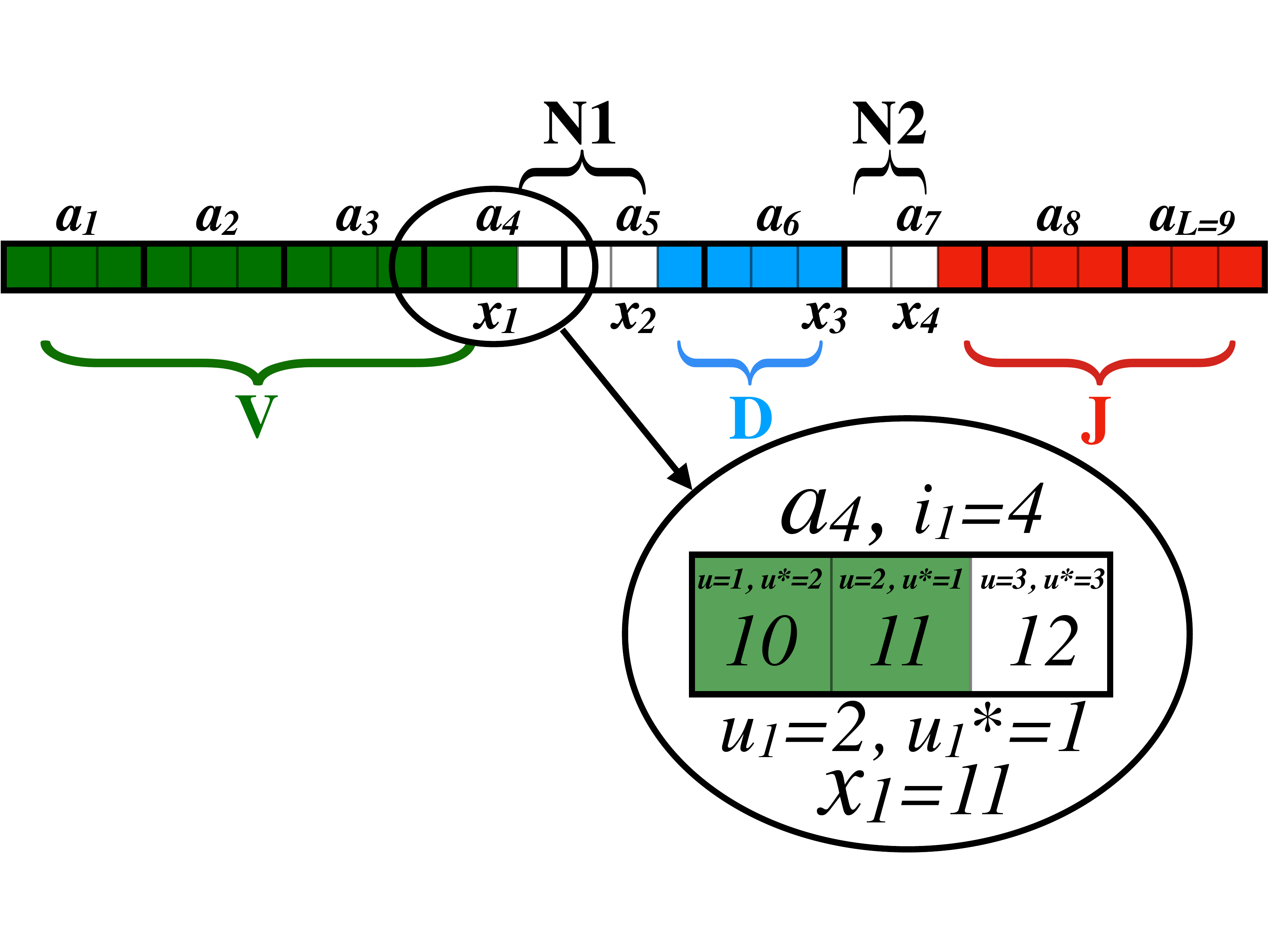}
\caption{\ZS{Schematic of the partitioning of an amino acid sequence into sections for the purpose of constructing the probability matrices underlying the dynamic programming method for computing its net generation probability. The indexing conventions are also highlighted.}\label{SI_cartoon}}
\end{center}
\end{figure}

The interpretation of the left (subscript) and right (superscript)
indices are detailed in the main text, \YE{and schematized in
  Fig.~\ref{SI_cartoon}. The sums are performed iteratively using
  matrix multiplications, as detailed in Fig.~\ref{olga_SI_schematic.pdf}.}
As in the main text, the
nucleotide indices will often be suppressed along with the implicit
dependence on the amino acid sequence $(a_1, \dots, a_L)$. For a given
nucleotide position $x_j$, it will be convenient to refer to the amino
acid index, and the position in the codon (from both the left and the
right), so we introduce the following (graphically shown in the cartoon below): $x_j=3(i_j-1)+u_j$, and $u$, so that $i_j$ encodes
the codon that index $x_j$ belongs to, and $u_j$ its position (from 1 to 3)
within that codon, while $u^*_j$ denotes the position taken from the right
of index $x_j+1$ within its codon, so that $u^*_j=2$ if
$u_j=1$, $u^*_j=1$ if $u_j=2$, and $u^*_j=3$ if $u_j=3$.

We now define the explicit forms for each of the matrices (note that we retain the indexing $x_j$ from Eq \ref{pgenaa_dyn_prog_sum}):

\subsubsection{$\mathcal{V}_{x_1}$}
Contribution from the templated V genes. $\mathcal{V}_{x_1}$ can be a 1x1 or 1x4 matrix depending on $u_1$. $\mathbf{s}^V$ is the sequence of the V germline gene (read \fivep to \threep) from the conserved residue (generally the cysteine C) to the end of the gene. $l_{V}$ is the length of $\mathbf{s}^V$. These equations are given in the main text.
\begin{equation}
\begin{split}
\mathcal{V}_{x_1}(\sigma)&=\sum_V P_{\rm V} (V)P_{\rm delV}(l_V-x_1|V) \mathbb{I}(s_{x_1}^V=\sigma)\mathbb{I}(\mathbf{s}^V_{1: {x_1}}\sim \mathbf{a}_{1: i_1})\quad\textrm{if }u_1=1,\\
\mathcal{V}_{x_1}(\sigma)&=\sum_V P_{\rm V} (V) P_{\rm delV}(l_V-{x_1}|V)\mathbb{I}((\mathbf{s}^V_{1: {x_1}},\sigma)\sim \mathbf{a}_{1: i_1})\quad\textrm{if }u_1=2,\\
\mathcal{V}_{x_1}&=\sum_V  P_{\rm V} (V) P_{\rm delV}(l_V-{x_1}|V)\mathbb{I}(\mathbf{s}^V_{1: {x_1}}\sim \mathbf{a}_{1: i_1})\quad\textrm{if }u_1=3.
\end{split}
\end{equation}

\subsubsection{${\mathcal{M}^{x_1}}_{x_2}$}
Contribution from the non-templated N1 insertions (VD junction). ${\mathcal{M}^{x_1}}_{x_2}$ is defined as the product of transfer matrices, and can be a 1x1, 1x4, 4x1, or 4x4 matrix depending on $u_1$ and $u_2$. The transfer matrices are defined by the summed contributions of the Markov insertion model of all codons consistent with the amino acid a (thus summations are over nucleotides $y,\  y_1,$ and $y_2$ to consider all allowed codons):
\begin{eqnarray}
T_{a}(\tau, \sigma) = \sum_{(y_1,y_2, \sigma) \sim a} S_{\rm VD}(\sigma | y_2)S_{\rm VD}(y_2 | y_1) S_{\rm VD}(y_1 | \tau)\\
F_{a}(\tau, \sigma) = S_{\rm VD}(\sigma | \tau)\mathbb{I}[\exists \sigma',\sigma''\textrm{ s.t. } (\sigma,\sigma',\sigma'')\sim a ]\\
D_{a} (\tau, \sigma) = \sum_{(y_1,y_2,\sigma) \sim a} S_{\rm VD}(y_2 | y_1) S_{\rm VD}(y_1 | \tau)\\
lT_{a}(\tau, \sigma) = \sum_{(\tau,y,\sigma) \sim a} S_{\rm VD}(\tau | y)p_0(y)\\
lD_{a}(\tau, \sigma) = \sum_{(\tau,y,\sigma) \sim a} p_0(y)
\end{eqnarray}

If $i_2 > i_1$:

\beq
{\mathcal{M}^{x_1}}_{x_2}= P_{\rm insVD}(x_2-x_1) L_{a_{i_1}}^{u_1} T_{a_{i_1+1}}\ldots T_{a_{i_2-1}} R_{a_{i_2}}^{u_2}
\eeq

where:
\beq
L_{a_{i_1}}^{u_1} =  \left\{
\begin{array}{cc} 
lT_{a_{i_1}} & \textrm{if } u_1 = 1\\
diag(p_0) & \textrm{if } u_1 = 2\\
S_{\rm VD}^{-1}p_0 & \textrm{if } u_1 = 3
\end{array}
\right.
\qquad {\rm and} \qquad
R_{a_{i_2}}^{u_2} = \left\{
\begin{array}{cc}
F_{a_{i_2}} & \textrm{if } u_2 = 1\\
D_{a_{i_2}} & \textrm{if } u_2 = 2\\
T_{a_{i_2}}\vec{1} & \textrm{if } u_2 = 3
\end{array}
\right.
\eeq

If $i_1 = i_2$:

\beq
{\mathcal{M}^{x_1}}_{x_2} = P_{\rm insVD}(x_2-x_1) \times \begin{array}{c|ccc}
 & u_2 = 1 & u_2 = 2 & u_2 = 3\\
\hline
u_1 = 1 & \mathds{1} & 0 & 0\\
u_1 = 2 & lD_{a_{i_1}} &  \mathds{1} & 0\\
u_1 = 3 & lT_{a_{i_1}}\vec{1} & diag(p_0)\vec{1} & 1
\end{array}
\eeq

\subsubsection{${{\mathcal{D}(D)}^{x_2}}_{x_3}$}
Contribution from the templated D genes. ${{\mathcal{D}(D)}^{x_2}}_{x_3}$ can be a 1x1, 1x4, 4x1, or 4x4 matrix depending on $u_2^*$ and $u_3^*$.  $\mathbf{s}^D$ is the sequence of the D germline gene (read \fivep to \threep) with length $l_D$.
\begin{equation}
\begin{split}
{\mathcal{D}(D)^{x_2}}_{x_3}(\tau, \sigma) &=\sum_{d^\prime_D} P_{\rm delD}(d_D, d^\prime_D|D)  \mathbb{I}[s^D_{d_D+1} = \tau] \mathbb{I}[s^D_{l_D - d^\prime_D} = \sigma] \mathbb{I}[\mathbf{s}^D_{d_D+1: l_D - d^\prime_D}\sim \mathbf{a}_{i_2: i_3}] \quad\textrm{if }u_2^*=1\textrm{ and } u_3^* = 1,\\
{\mathcal{D}(D)^{x_2}}_{x_3}(\tau, \sigma) &=\sum_{d^\prime_D} P_{\rm delD}(d_D, d^\prime_D|D)  \mathbb{I}[s^D_{dD+1} = \tau]\mathbb{I}[(\mathbf{s}^D_{d_D + 1: l_D - d^\prime_D}, \sigma)\sim \mathbf{a}_{i_2: i_3}] \quad\textrm{if }u_2^*=1\textrm{ and } u_3^* = 2,\\
{\mathcal{D}(D)^{x_2}}_{x_3}(\tau) &=\sum_{d^\prime_D} P_{\rm delD}(d_D, d^\prime_D|D)  \mathbb{I}[s^D_{d_D+1} = \tau] \mathbb{I}[\mathbf{s}^D_{d_D+1: l_D - d^\prime_D}\sim \mathbf{a}_{i_2: i_3}] \quad\textrm{if }u_2^*=1\textrm{ and } u_3^* = 3,
\end{split}
\end{equation}
\begin{equation}
\begin{split}
{\mathcal{D}(D)^{x_2}}_{x_3}(\tau, \sigma) &=\sum_{d^\prime_D} P_{\rm delD}(d_D, d^\prime_D|D) \mathbb{I}[s^D_{l_D - d^\prime_D} = \sigma] \mathbb{I}[(\tau, \mathbf{s}^D_{d_D+1: l_D - d^\prime_D})\sim \mathbf{a}_{i_2: i_3}] \quad\textrm{if }u_2^*=2\textrm{ and } u_3^* = 1,\\
{\mathcal{D}(D)^{x_2}}_{x_3}(\tau, \sigma) &=\sum_{d^\prime_D} P_{\rm delD}(d_D, d^\prime_D|D) \mathbb{I}[(\tau, \mathbf{s}^D_{d_D+1: l_D - d^\prime_D}, \sigma)\sim \mathbf{a}_{i_2: i_3}] \quad\textrm{if }u_2^*=2\textrm{ and } u_3^* = 2,\\
{\mathcal{D}(D)^{x_2}}_{x_3}(\tau) &=\sum_{d^\prime_D} P_{\rm
  delD}(d_D, d^\prime_D|D) \mathbb{I}[(\tau, \mathbf{s}^D_{d_D+1: l_D
  - d^\prime_D})\sim \mathbf{a}_{i_2: i_3}] \quad\textrm{if
}u_2^*=2\textrm{ and } u_3^* = 3,
\end{split}
\end{equation}
\begin{equation}
\begin{split}
{\mathcal{D}(D)^{x_2}}_{x_3}(\sigma) &=\sum_{d^\prime_D} P_{\rm delD}(d_D, d^\prime_D|D) \mathbb{I}[s^D_{l_D - d^\prime_D} = \sigma] \mathbb{I}[\mathbf{s}^D_{d_D+1: l_D - d^\prime_D}\sim \mathbf{a}_{i_2: i_3}] \quad\textrm{if }u_2^*=3\textrm{ and } u_3^* = 1,\\
{\mathcal{D}(D)^{x_2}}_{x_3}(\sigma) &=\sum_{d^\prime_D} P_{\rm delD}(d_D, d^\prime_D|D)  \mathbb{I}[(\mathbf{s}^D_{d_D+1: l_D - d^\prime_D}, \sigma)\sim \mathbf{a}_{i_2: i_3}] \quad\textrm{if }u_2^*=3\textrm{ and } u_3^* = 2,\\
{\mathcal{D}(D)^{x_2}}_{x_3} &=\sum_{d^\prime_D} P_{\rm delD}(d_D, d^\prime_D|D) \mathbb{I}[\mathbf{s}^D_{d_D+1: l_D - d^\prime_D}\sim \mathbf{a}_{i_2: i_3}] \quad\textrm{if }u_2^*=3\textrm{ and } u_3^* = 3
\end{split}
\end{equation}
where $d_D = l_D - (x_3-x_2) - d^\prime_D$ \\

\subsubsection{${\mathcal{N}^{x_3}}_{x_4}$}
Contribution from the non-templated N2 insertions (DJ junction). ${\mathcal{N}^{x_3}}_{x_4}$ is defined as the product of transfer matrices, and can be a 1x1, 1x4, 4x1, or 4x4 matrix depending on $u_3^*$ and $u_4^*$. The transfer matrices are defined by the summed contributions of the Markov insertion model of all codons consistent with the amino acid a (thus summations are over nucleotides $y,\  y_1,$ and $y_2$ to consider all allowed codons):
\begin{eqnarray}
T^\prime_{a}(\tau, \sigma) = \sum_{(\sigma, y_2, y_1) \sim a} S_{\rm DJ}(\sigma | y_2)S_{\rm DJ}(y_2 | y_1) S_{\rm DJ}(y_1 | \tau)\\
F^\prime_{a}(\tau, \sigma) = S_{\rm DJ}(\sigma | \tau)\mathbb{I}[\exists \sigma',\sigma''\textrm{ s.t. } (\sigma'',\sigma', \sigma)\sim a ]\\
D^\prime_{a} (\tau, \sigma) = \sum_{(\sigma, y_2, y_1) \sim a} S_{\rm DJ}(y_2 | y_1) S_{\rm DJ}(y_1 | \tau)\\
lT^\prime_{a}(\tau, \sigma) = \sum_{(\sigma, y, \tau) \sim a} S_{\rm DJ}(\tau | y)q_0(y)\\
lD^\prime_{a}(\tau, \sigma) = \sum_{(\sigma, y, \tau) \sim a} q_0(y)
\end{eqnarray}

If $i_4 > i_3$:

\beq\
{\mathcal{N}^{x_3}}_{x_4}= P_{\rm insDJ}(x_4-x_3) {L^\prime}_{a_{i_3}}^{u_3^*} T^\prime_{a_{i_3+1}}\ldots T^\prime_{a_{i_4-1}} {R^\prime}_{a_{i_4}}^{u_4^*}
\eeq

where:
\beq
{L^\prime}_{a_{i_3}}^{u_3^*} = \left\{
\begin{array}{cc}
F^\prime_{a_{i_3}} & \textrm{if } u_3^* = 1\\
D^\prime_{a_{i_3}} & \textrm{if } u_3^* = 2\\
T^\prime_{a_{i_3}}\vec{1} & \textrm{if } u_3^* = 3
\end{array}
\right.
\qquad {\rm and} \qquad
{R^\prime}_{a_{i_4}}^{u_4^*} =  \left\{
\begin{array}{cc} 
lT^\prime_{a_{i_4}} & \textrm{if } u_4^* = 1\\
diag(q_0) & \textrm{if } u_4^* = 2\\
S_{\rm DJ}^{-1}q_0 & \textrm{if } u_4^* = 3
\end{array}
\right.
\eeq

If $i_3 = i_4$:

\beq
{\mathcal{N}^{x_3}}_{x_4} = P_{\rm insDJ}(x_4-x_3) \times \begin{array}{c|ccc}
 & u_4^* = 1 & u_4^* = 2 & u_4^* = 3\\
\hline
u_3^* = 1 & \mathds{1} & lD^\prime_{a_{i_3}} & lT^\prime_{a_{i_3}}\vec{1}\\
u_3^* = 2 & 0 &  \mathds{1} &  diag(q_0)\vec{1}\\
u_3^* = 3 & 0 & 0 & 1
\end{array}
\eeq

\subsubsection{${\mathcal{J}(D)}^{x_4}$}
Contribution from the templated J genes. ${\mathcal{J}(D)}^{x_4}$ can be a 1x1 or 4x1 matrix depending on $u_4^*$. $\mathbf{s}^J$ is the sequence of the J germline gene (read \fivep to \threep) and $l_J$ gives the length of the sequence up to the conserved residue (generally either F or W).
\begin{equation}
\begin{split}
\mathcal{J}(D)^{x_4}(\tau)&=\sum_J  P_{\rm D, J} (DJ) P_{\rm delJ}(d_J|J) \mathbb{I}(s^J_{d_J+1}=\tau)\mathbb{I}(\mathbf{s}^J_{d_J+1: l_J}\sim \mathbf{a}_{i_4: L})\quad\textrm{if }u_4^*=1,\\
\mathcal{J}(D)^{x_4}(\tau)&=\sum_J  P_{\rm D, J} (DJ) P_{\rm delJ}(d_J|J)\mathbb{I}((\tau, \mathbf{s}^J_{d_J+1: l_J})\sim \mathbf{a}_{i_4: L})\quad\textrm{if } u_4^*=2,\\
\mathcal{J}(D)^{x_4}&=\sum_J  P_{\rm DJ} (D, J) P_{\rm delJ}(d_J|J)\mathbb{I}(\mathbf{s}^J_{d_J+1: l_J})\sim \mathbf{a}_{i_4: L})\quad\textrm{if } u_4^*=3.
\end{split}
\end{equation}
where $dJ = l_J - 3L - x_4 - 1$\\

\begin{figure}[th!]
\centering
\includegraphics[width=.7\linewidth]{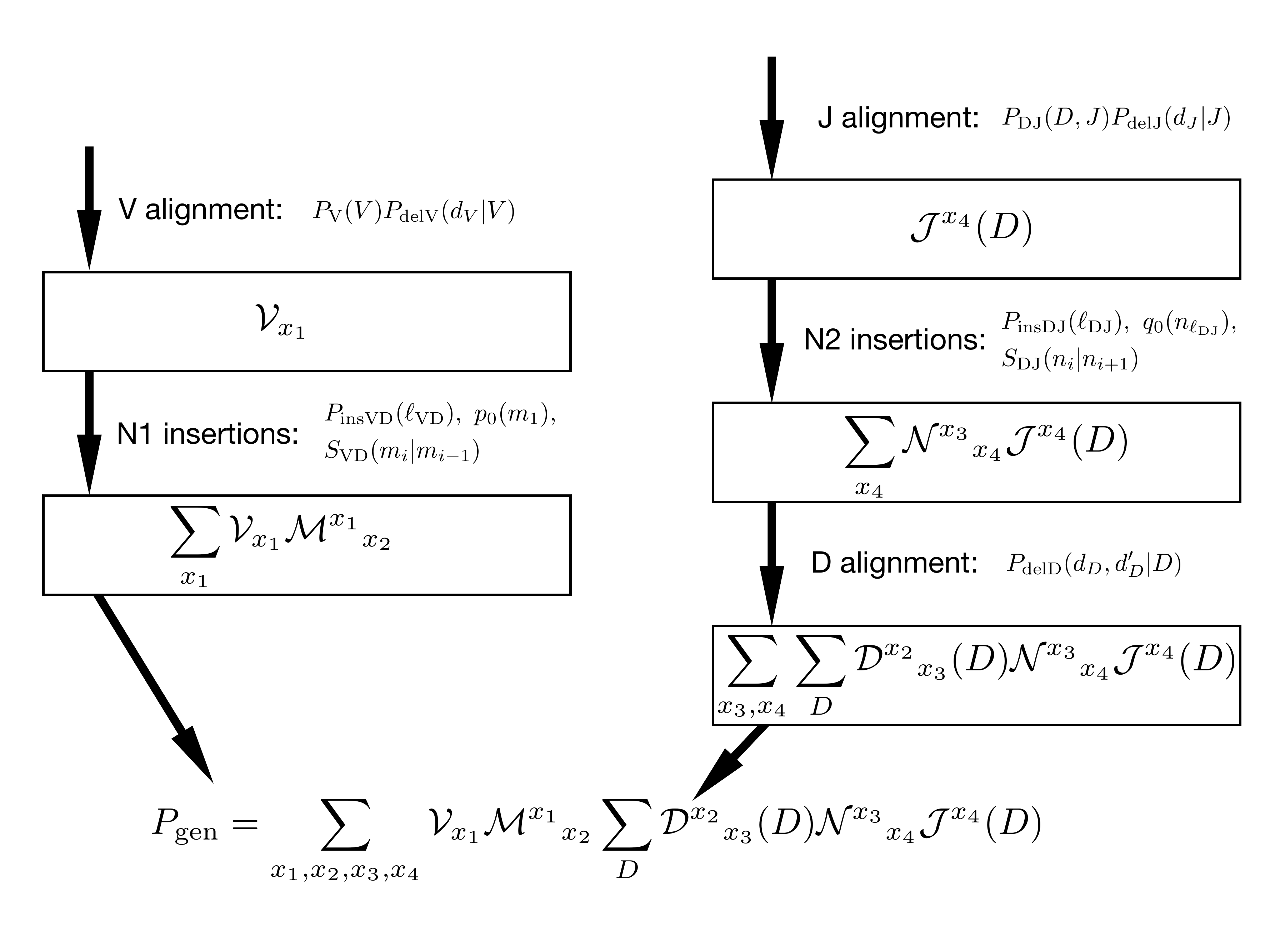}
\caption{\ZS{Schematic of the OLGA VDJ algorithm implementation breakdown. Each of the 5 segments (V, N1, D, N2, J), and their associated model contributions, are considered from the edges of the CDR3 towards the inside. This is done both from the left side (V, N1) and the right side (D, N2, J) of the read to efficiently account for the correlations for the D and J genes. Including inner segments (N2, D, N2) requires summing over an index, indicating that all possible allowed start and end positions of the segment are considered.}\label{SI_schematic}}
\end{figure}

\section{VJ recombination}

The model used for VJ recombination is quite similar to the model for VDJ recombination with the main differences being the lack of a D segment and an N2 insertion segment. However, a strong correlation between V and J templates is observed in the TRA chain, so we include a joint V, J distribution to allow for this correlation. Due to this similarity, the algorithm used to compute $\rm P_{gen}$ is very similar. The VJ generative model is:

\beq\label{alpha_genmodel}
P^{\rm rec}_{\rm gen}(E) = P_{\rm VJ}(V, J)P_{\rm delV}(d_V | V) P_{\rm delJ}(d_J|J) \times P_{\rm insVJ}(\ell_{\rm VJ})p_0(m_1)\left[\prod_{i=2}^{\ell_{VJ}} S_{\rm VJ}(m_{i}|m_{i-1})\right]
\eeq

with nucleotide and amino acid $P_{\rm gen}$s being defined the same as for the VDJ recombination model (Eq \ref{pgenaa_sum_over_events}). The dynamic programing algorithm also has a similar form to Eq \ref{pgenaa_dyn_prog_sum}, and can be summarized as (retaining all notation conventions from before):
\begin{equation}\label{pgenmatrix}
P_{\rm gen}(a_1,\ldots,a_L)=\sum_{x_1,x_2}\sum_{J}{\mathcal{V}(J)}_{x_1}{\mathcal{M}^{x_1}}_{x_2} {\mathcal{J}(J)}^{x_2}
\end{equation}

\subsubsection{$\mathcal{V(J)}_{x_1}$}
Contribution from the templated V genes.
\begin{equation}
\begin{split}
{\mathcal{V}(J)}_{x_1}(\sigma)&=\sum_V P_{\rm VJ} (V, J)P_{\rm delV}(l_V-x_1|V) \mathbb{I}(s_{x_1}^V=\sigma)\mathbb{I}(\mathbf{s}^V_{1: {x_1}}\sim \mathbf{a}_{1: i_1})\quad\textrm{if }u_1=1,\\
{\mathcal{V}(J)}_{x_1}(\sigma)&=\sum_V P_{\rm VJ} (V, J) P_{\rm delV}(l_V-{x_1}|V)\mathbb{I}((\mathbf{s}^V_{1: {x_1}},\sigma)\sim \mathbf{a}_{1: i_1})\quad\textrm{if }u_1=2,\\
{\mathcal{V}(J)}_{x_1}&=\sum_V  P_{\rm VJ} (V, J) P_{\rm delV}(l_V-{x_1}|V)\mathbb{I}(\mathbf{s}^V_{1: {x_1}}\sim \mathbf{a}_{1: i_1})\quad\textrm{if }u_1=3.
\end{split}
\end{equation}

\subsubsection{${\mathcal{M}^{x_1}}_{x_2}$}
Contribution from the non-templated N insertions (VJ junction). ${\mathcal{M}^{x_1}}_{x_2}$ is identical to the definition of ${\mathcal{M}^{x_1}}_{x_2}$ from the VDJ algorithm (except using the parameters $S_{\rm VJ}, P_{\rm insVJ},$ and $p_0$ from a VJ recombination model).

\subsubsection{${\mathcal{J}(J)}^{x_2}$}
Contribution from the templated J genes.
\begin{equation}
\begin{split}
\mathcal{J}(J)^{x_2}(\tau)&= P_{\rm delJ}(d_J|J) \mathbb{I}(s^J_{d_J+1}=\tau)\mathbb{I}(\mathbf{s}^J_{d_J+1: l_J}\sim \mathbf{a}_{i_2: L})\quad\textrm{if }u_2^*=1,\\
\mathcal{J}(J)^{x_2}(\tau)&= P_{\rm delJ}(d_J|J)\mathbb{I}((\tau, \mathbf{s}^J_{d_J+1: l_J})\sim \mathbf{a}_{i_2: L})\quad\textrm{if } u_2^*=2,\\
\mathcal{J}(J)^{x_2}&= P_{\rm delJ}(d_J|J)\mathbb{I}(\mathbf{s}^J_{d_J+1: l_J})\sim \mathbf{a}_{i_2: L})\quad\textrm{if } u_2^*=3.
\end{split}
\end{equation}
where $dJ = l_J - 3L - x - 1$\\

\begin{figure}
\centering
\includegraphics[width=.5\linewidth]{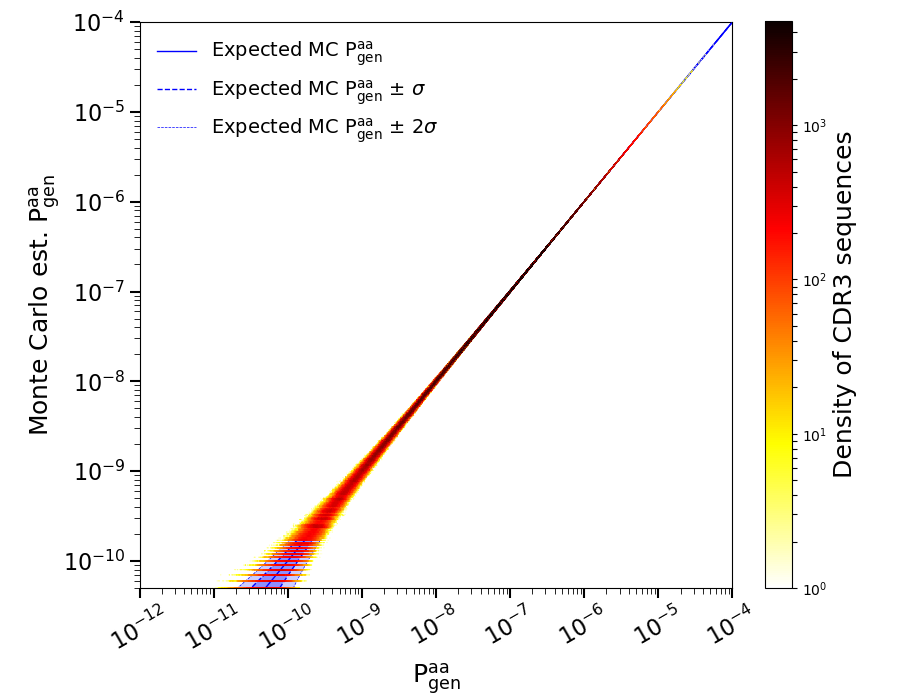}
\caption{\ZS{Monte Carlo estimate of the generation probability of amino acid human TRA CDR3 sequences, $P_{\rm gen}^{\rm aa}$ , versus OLGA's calculation. The horizontal lines at the lower left of the plot represent CDR3s that were generated once, twice, etc, in the MC sample. The one- and two-sigma curves display the deviations from exact equality between simulated and computed $P_{\rm gen}$ to be expected on the basis of Poisson statistics.\label{alpha_chain_validation}}}
\end{figure}

This algorithm is validated in the same manner to the VDJ algorithm, i.e. comparing to Monte Carlo (MC) estimation (\ZS{Fig \ref{alpha_chain_validation}}).

\section{Dependence on model parameters and structure}
\ZS{In order to efficiently compute the summation in Eq. \ref{pgenaa_dyn_prog_sum} the summations of the model contributions from each of the 5 segments of a CDR3 (V genomic, N1 insertions, D genomic, N2 insertions, and J genomic) are performed in a specific order (summarized in Fig \ref{SI_schematic}). Specifically, we start at the left and right ends of the CDR3 read and move inwards, summing over positional indices at each step. As the D and J segments are correlated, it is useful to consider the V and N1 contributions separately from the D, N2, and J and to do the final summation over the index $x_2$ after the D, N2, and J components are summed over all D alleles (notice the D dependencies in Fig \ref{SI_schematic}). }
\ZS{This breakdown is useful to highlight the most computationally intensive steps: N2 insertions and the D alignment. These steps (along with the N1 insertions) require considering that the associated segment could begin and end at each allowed position. This is mathematically seen as the summation over positions and computing a matrix indexed by two indices, leading to an $O(L^2)$ complexity. The N2 insertions and D alignments are further aggravated due to model correlations between the D and J genes requiring repeating the steps for N2 insertions and D alignment for each D allele. The runtime of OLGA is thus most sensitive to the maximum number of N2 insertions and the length and number of the D alleles. The effects of varying these parameters is best illustrated by comparing runtimes for mouse TRB, human TRB, and human IGH models (Table \ref{tab:model_comp}). }
\begin{table}[h!]
  \centering
  \caption{Model comparison}
  \label{tab:model_comp}
  \begin{tabular}{c|c|c|c}
    \toprule
    Species/Chain & max insertions & $\#$ D alleles & Average computation speed \\
    \hline
    Mouse TRB & 11 & 2 & 70.4 seqs/CPU second \\
    Human TRB & 30 & 3 & 35.6 seqs/CPU second \\
    Human IGH & 60 & 35 & 2.05 seqs/CPU second \\
    \bottomrule
  \end{tabular}
\end{table}
\ZS{In a similar fashion, the most computationally intensive step of computing $P_{\rm gen}$ of a VJ model (e.g. human TRA) is the insertion step, and due to correlations between the V and J genes this is repeated for each J allele in a similar fashion as the D alleles However, as the J region of a human TRA is fairly large, many of these J genes can be excluded from alignment (if they contribute 0 probability), yielding the much faster computation rate of 184 seqs/CPU second.}

\section{Timing, performance, model dependence}
\ZS{In order to analyze OLGA's computational performance as a function of CDR3 length, and to compare to other hypothetical methods, we use the human TRB model as an example.}

\begin{figure}
\begin{center}
\includegraphics[width=1\linewidth]{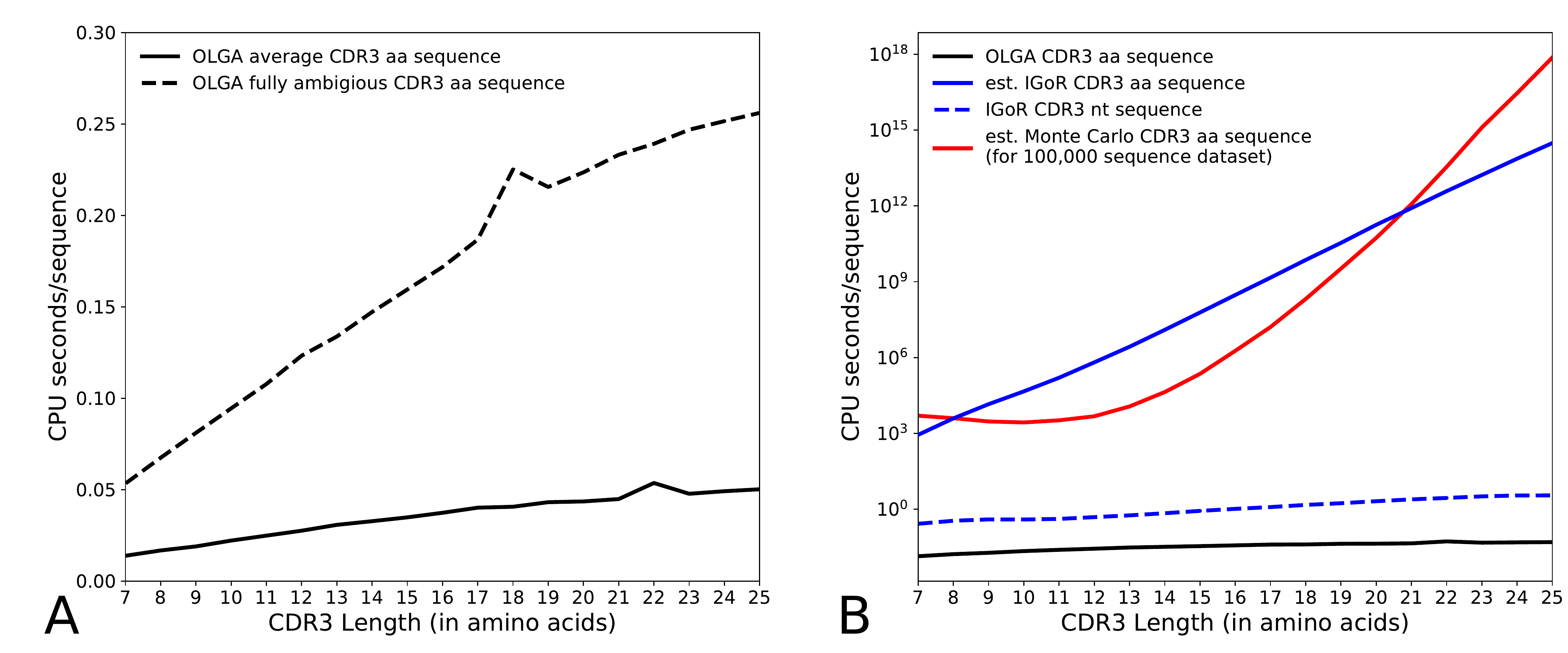}
\caption{\ZS{A) Computational performance of OLGA as a function of CDR3 length. We compare performance averaged over a sample of human TRB amino acid CDR3 sequences to the worst case scenario of CDR3 sequences composed of fully ambiguous amino acids X. In both cases the time for a single sequence increases roughly linearly (i.e. less than the algorithmic worst case of $O(L^2)$). B) Computational performance of different $P_{\rm gen}$ methods as a function of CDR3 length (log scale). The IGoR and OLGA runtimes are determined by running over the same statistical sample of human TRB sequences. OLGA runs over the translated amino acid CDR3 sequences while IGoR runs over nucleotide CDR3 (dashed blue line). In order to compare OLGA to how long it would take IGoR to compute $P_{\rm gen}$ of amino acid CDR3s we estimate by multiplying the IGoR runtime of single nucleotide sequences (dashed blue line) by the number of nucleotide sequences that translate to the given amino acid sequence (yielding the solid blue line). Monte Carlo runtime is estimated for a dataset of 100,000 sequences with an estimated coverage of 66$\%$ of sequences having at least one count. OLGA vastly outperforms both direct enumeration (est. IGoR) and Monte Carlo.}\label{timing_by_cdr3_length}}
\end{center}
\end{figure}

\ZS{As discussed in the previous section, the most computationally intensive steps of OLGA (N1, N2, and D) require at most $O(L^2)$ operations. In practice, OLGA's scaling of the computation speed as a function of CDR3 length, even for the worst case sequences, i.e. fully ambiguous amino acids of a given length, is closer to linear in the relevant regime due to the finite parameterization of the model (maximum number of insertions, maximum size of D sequences, etc). This is shown in Fig \ref{timing_by_cdr3_length}A.}

\ZS{We also compare OLGA to runtimes of IGoR (i.e. direct enumeration of recombination events) and a hypothetical Monte Carlo computation (Fig \ref{timing_by_cdr3_length}). As we will explain, neither the IGoR nor the MC are precise comparisons to OLGA, yet OLGA is faster than either. }

\ZS{The IGoR runtimes are for nucleotide sequences not amino acid sequences. In order for IGoR to compute the $P_{\rm gen}$ of an amino acid sequence, it would need to compute and sum the $P_{\rm gen}$ of each nucleotide sequence that codes for the amino acid sequence. These sequences can be enumerated for extremely short CDR3 lengths, however the number explodes exponentially in CDR3 length. Even for a CDR3 length of 4, by enumerating all nucleotide sequences for an amino acid sequence IGoR computes 0.33 seqs/CPU second compared to the 122 seqs/CPU second for OLGA. For longer CDR3 lengths we approximate how long IGoR would take by computing the average number of CDR3 nucleotide sequences per CDR3 amino acid sequence for a given length. OLGA not only heavily outperforms this exponential blowup, but actually outperforms IGoR when IGoR is computing a $single$ nucleotide sequence of a given amino acid sequence.}

\begin{figure}
\begin{center}
\includegraphics[width=1\linewidth]{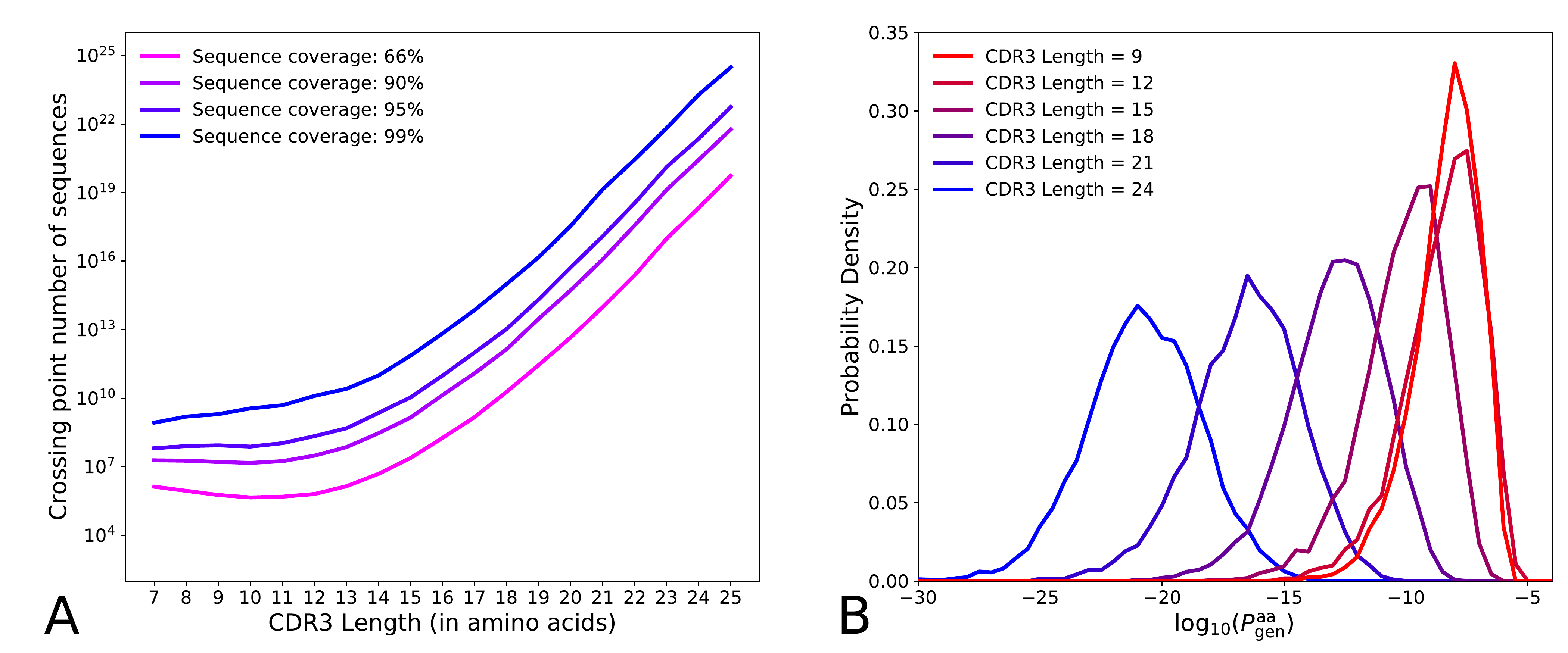}
\caption{\ZS{A) The runtime of Mont Carlo $P_{\rm gen}$ estimation scales as ${1}/{P_{\rm gen}}$ while OLGA will scale with the number of sequences. This predicts a number of sequences for the \lq crossing point\rq\ where the runtime of Monte Carlo $P_{\rm gen}$ estimation is comparable to OLGA $P_{\rm gen}$ for sequences with $P_{\rm gen}$ above some cutoff. For datasets with more sequences than these curves, Monte Carlo estimation may be faster (depending on the level of Poisson noise considered tolerable), while below these curves OLGA is always faster. We plot this as a function of CDR3 length where the $P_{\rm gen}$ cutoffs are determined to ensure that on average some fraction (66$\%$, 90$\%$, 95$\%$, and 99$\%$) of the sequences at that length get covered by the Monte Carlo estimation. B) $\log_{10}(P_{\rm gen})$ probability density distributions for a few examples of CDR3 lengths. These curves are used to determine the MC $P_{\rm gen}$ cutoffs per CDR3 length by determining, for a given curve, when the area under the curve and right of a $P_{\rm gen}$ cutoff matches the sequence coverage fraction.}\label{pgen_hists_by_cdr3_length_and_mc_cp}}
\end{center}
\end{figure}

\ZS{The Monte Carlo runtime estimate comes from the setup of estimating the $P_{\rm gen}$ of 100,000 sequences. These $P_{\rm gen}$ would be estimated by simulating enough recombination events such that 66$\%$ of CDR3 sequences of a given length would be expected to have at least one count. There is a CDR3 length scaling due to the trend that shorter sequences tend to have higher $P_{\rm gen}$ (Fig \ref{pgen_hists_by_cdr3_length_and_mc_cp}B). The $P_{\rm gen}$ estimated using this methodology will be extremely noisy (Poisson noise on the expected number of counts) and not even give reliable estimates for many sequences.}

\ZS{It is true that the computation time for MC estimates scale as ${1}/{P_{\rm gen}}$ and not with the number of sequences. Thus, there is a hypothetical number of sequences when MC is faster than OLGA if we are willing to accept noisy estimates and to entirely miss some fraction of the CDR3s. This \lq crossing point\rq\ number of sequences is plotted in Fig \ref{pgen_hists_by_cdr3_length_and_mc_cp}A and corresponds to completely unrealistic numbers of sequences, highlighting the fact that OLGA will not only give a more reliable $P_{\rm gen}$, even for very unlikely sequences, but is also much faster than MC even for short, high $P_{\rm gen}$, sequences. So, even overlooking the drawbacks and imprecision of MC estimation, for plausible sized datasets OLGA is still dramatically faster than MC.}

\section{Generation probability distributions from RNA-derived repertoires}

\YE{The analyses described in the main text were mostly concerned with datasets derived by sequencing the genomic DNA contained in a sample of immune cells to directly obtain sequences of the rearranged TCR genes. Immune repertoires can alternatively be obtained by sequencing the mRNA expressed from the same genes, and many such RNA-based data sets exist. Given a TCR sequence, OLGA evaluates the probability of the primitive recombination event (or events) that must have occurred to create the initial T cell carrying that sequence, and the applicability of OLGA is independent of how the sequence was obtained (i.e. from DNA or RNA sequencing). OLGA relies on the availability of a suitable recombination model but that model is thought to vary very little with time (and disease status) for each individual subject and only moderately from individual to individual in a given species. The probability that a given sequence, once generated in a primitive event, will be captured in a sequencing experiment is at best roughly constant across sequences, and may vary substantially between different capture protocols.}

\begin{figure}
\centering
\includegraphics[width=.6\linewidth]{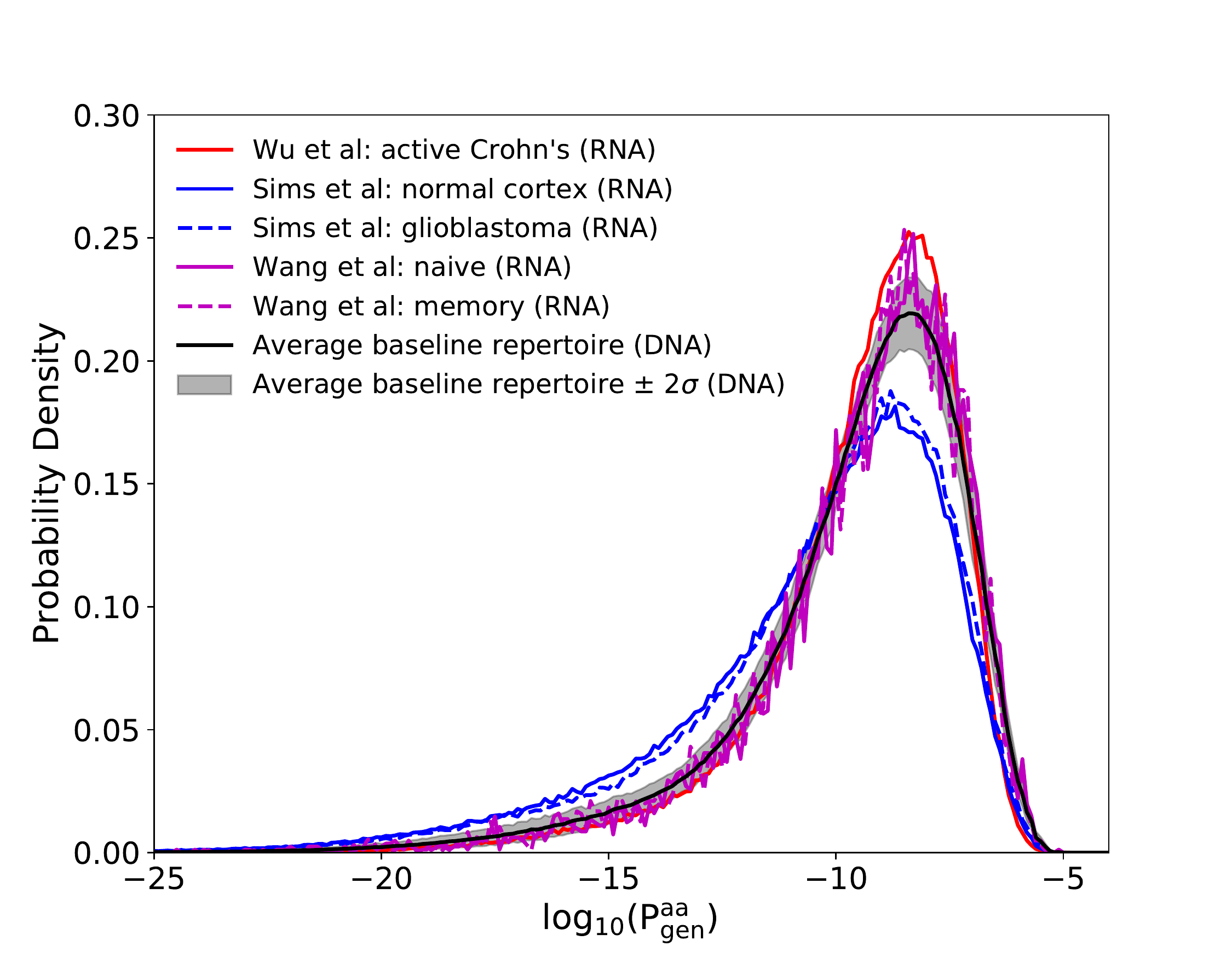}
\caption{\ZS{Generation probability distributions for TRB CDR3 sequences taken from three different sources (\citet{Sims2016}, \citet{Wu2018}, \citet{Wang2010}), compared to DNA RepSeq data from \citet{Emerson2017} (black curve with standard deviation), using a model inferred from \citep{Emerson2017}. All distributions have approximately the same shape, with a slight bias in the data from \citet{Sims2016}, indicating how robust is the distribution. Data from Sims {\em et al.} are identified in the SI of \cite{Sims2016} as IDs N01 (normal cortex) and G10 (glioblastoma). Data from Wang {\em et al.} is identified by Short Read Archive (SRA) accession numbers SRR030702 (naive) and SRS007450 (memory). See also Fig.~\ref{Sims}.}\label{pgen_dist_sources}}
\end{figure}

\begin{figure}
\centering
\includegraphics[width=.6\linewidth]{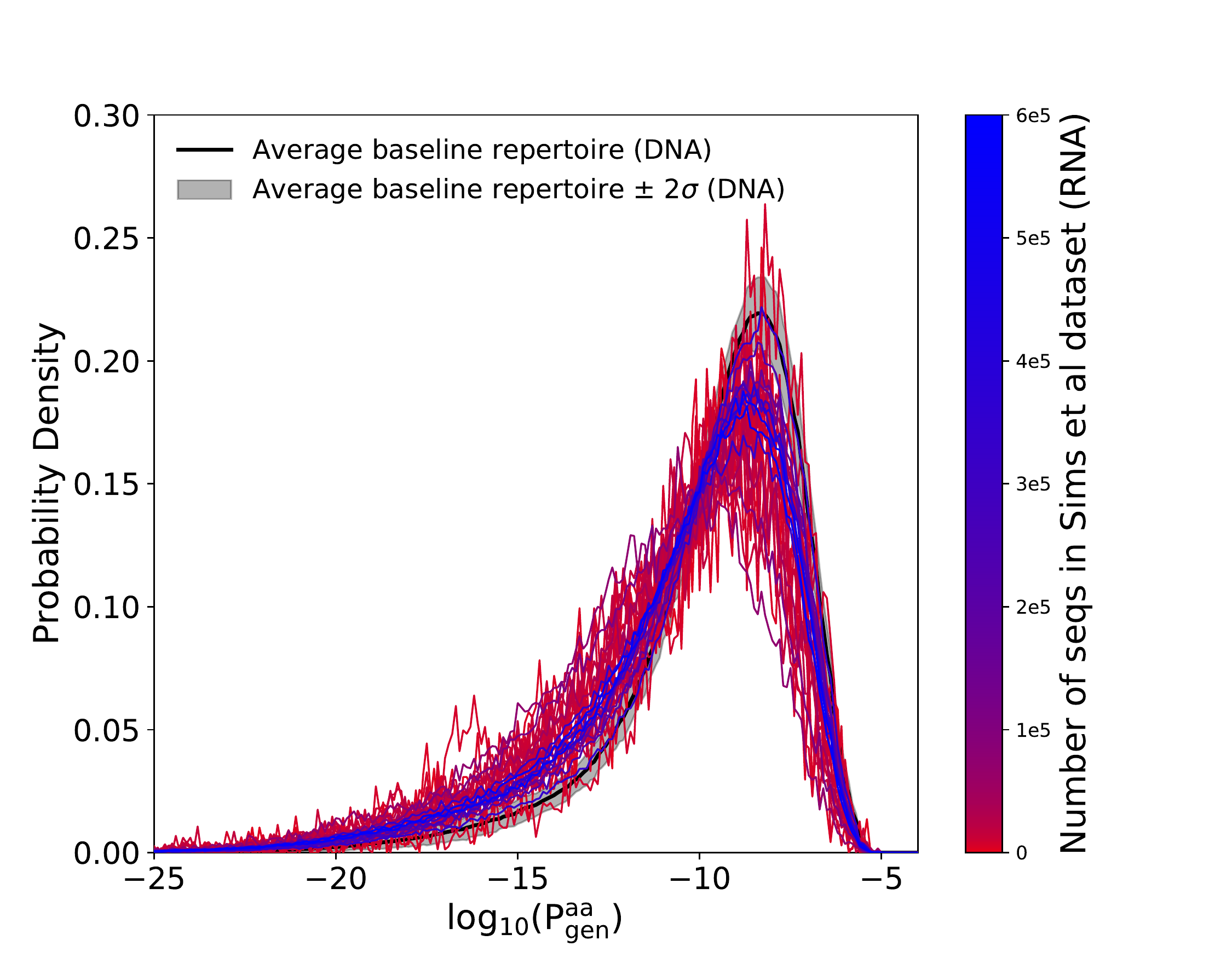}
\caption{\ZS{Generation probability distributions for different samples from \citet{Sims2016}, compared to DNA RepSeq data from \citet{Emerson2017} (black curve with standard deviation), using a model inferred from \citep{Emerson2017}. Color indicate sample size: larger datasets are blue, while small ones are red. The only effect of decreasing the sample size is increasing the noise, but the shape stays the same. All TRB datasets from the study are plotted.}\label{Sims}}
\end{figure}

\YE{For these reasons, it is interesting to investigate how these generation probability distributions vary across CDR3 repertoires obtained using different sequencing protocols in different biological contexts. In Fig.\,\ref{pgen_dist_sources} we plot the results of running OLGA on a few recently published human TRB repertoires that were obtained using RNA sequencing. These samples comprise a study of patients with glioblastoma disease (\citet{Sims2016}), a study of patients with Crohn's disease and ulcerative colitis (\citet{Wu2018}), and a comprehensive study of the dynamics of TCRs in healthy individuals (\citet{Wang2010}). Fig.\,\ref{pgen_dist_sources} shows the generation probability distribution of data sets from these three sources, for comparison plotted together with the distribution obtained from DNA sequencing of the large human sample of \citet{Emerson2017}.  As can be seen, two of the three RNA data sets give results quite consistent with the DNA-based results. The glioblastoma data (\citet{Sims2016}) gives a distribution broadly similar to the other three, but with a systematic shift to higher frequency of occurrence of lower generation probability sequences. We do not know whether or not this difference is biologically significant, or an artifact of the used protocol. The difference does not seem to be due to sampling depth, as can be seen in Fig.\,\ref{Sims}, where multiple samples from \citet{Sims2016} are plotted: the distributions derived from smaller samples are noisier than, but statistically consistent with, the distributions based on the largest samples.
}

\section{Cross-species $P_{\rm gen}$}
\begin{figure}
\centering
\includegraphics[width=.7\linewidth]{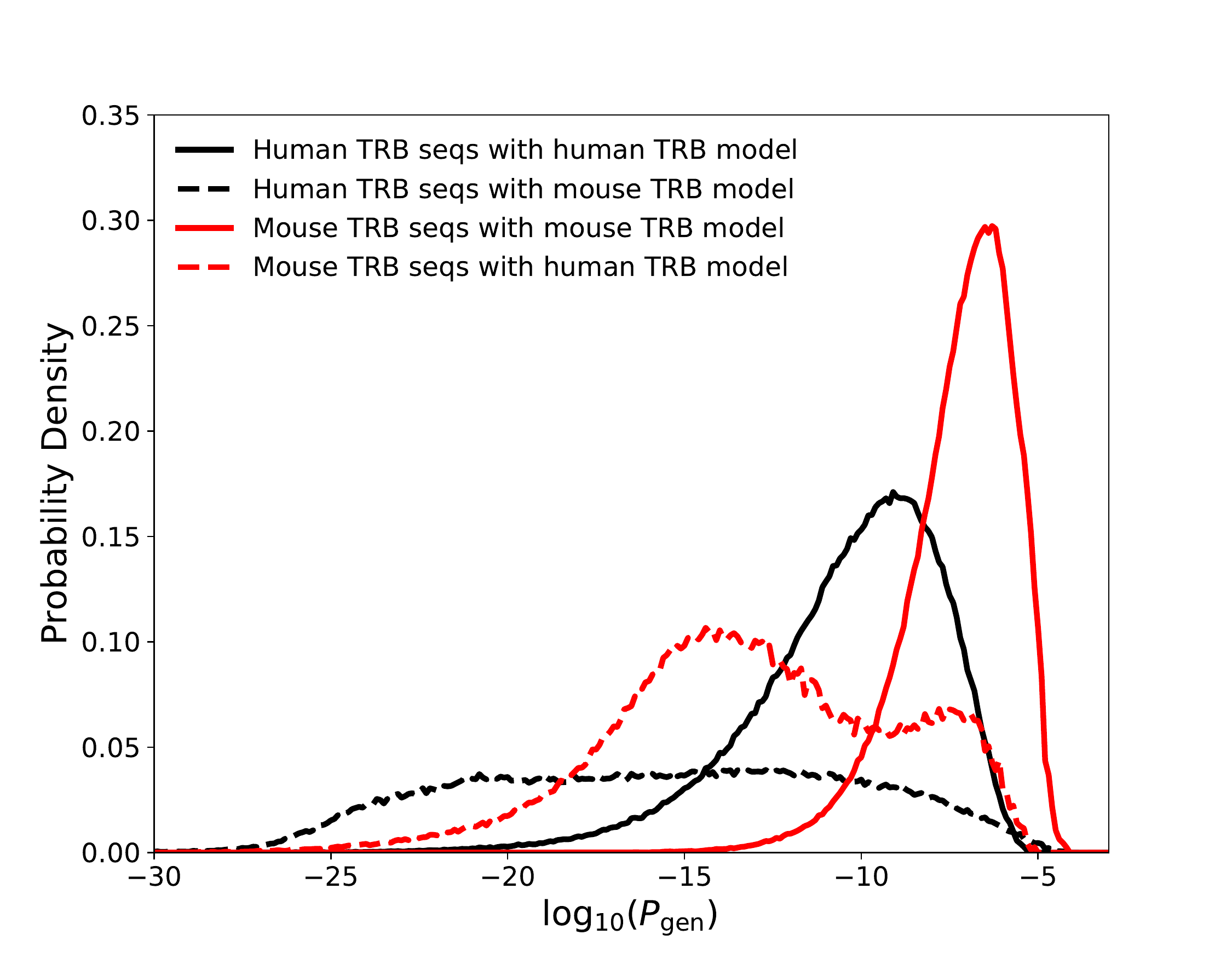}
\caption{\ZS{Probability densities of $\log_{10}(P_{\rm gen})$ for sequences generated from mouse TRB and human TRB models. The $P_{\rm gen}$ of a sequence is computed using either a mouse TRB model or a human TRB model depending on the curve. Models are based on data from \citet{Emerson2017} for human TRB and \citet{Sethna2017} for mouse TRB.}
\label{human_mouse_hist}}
\end{figure}

\YE{TCR sequence repertoires are different in detail between species, both because the genomic templates differ and because of differences in the parameters of the recombination process itself. As a result, there are clear interspecies differences in CDR3 length distribution and amino acid composition. Nevertheless, the TRB CDR3 regions of different vertebrate species have the same overall structure and the same conserved residues at the two ends of the CDR3. As a result, a CDR3 from one species usually has a non-zero probability to be produced within a different species, a fact of some interest in the context of studies of cross-species sharing of T cell types. We explored this concept with OLGA by feeding TRB CDR3s produced by the human generation model to a mouse generation model and vice versa. The resulting generation probability distributions are plotted in Fig.\,\ref{human_mouse_hist}.}

\begin{figure}
\centering
\includegraphics[width=1\linewidth]{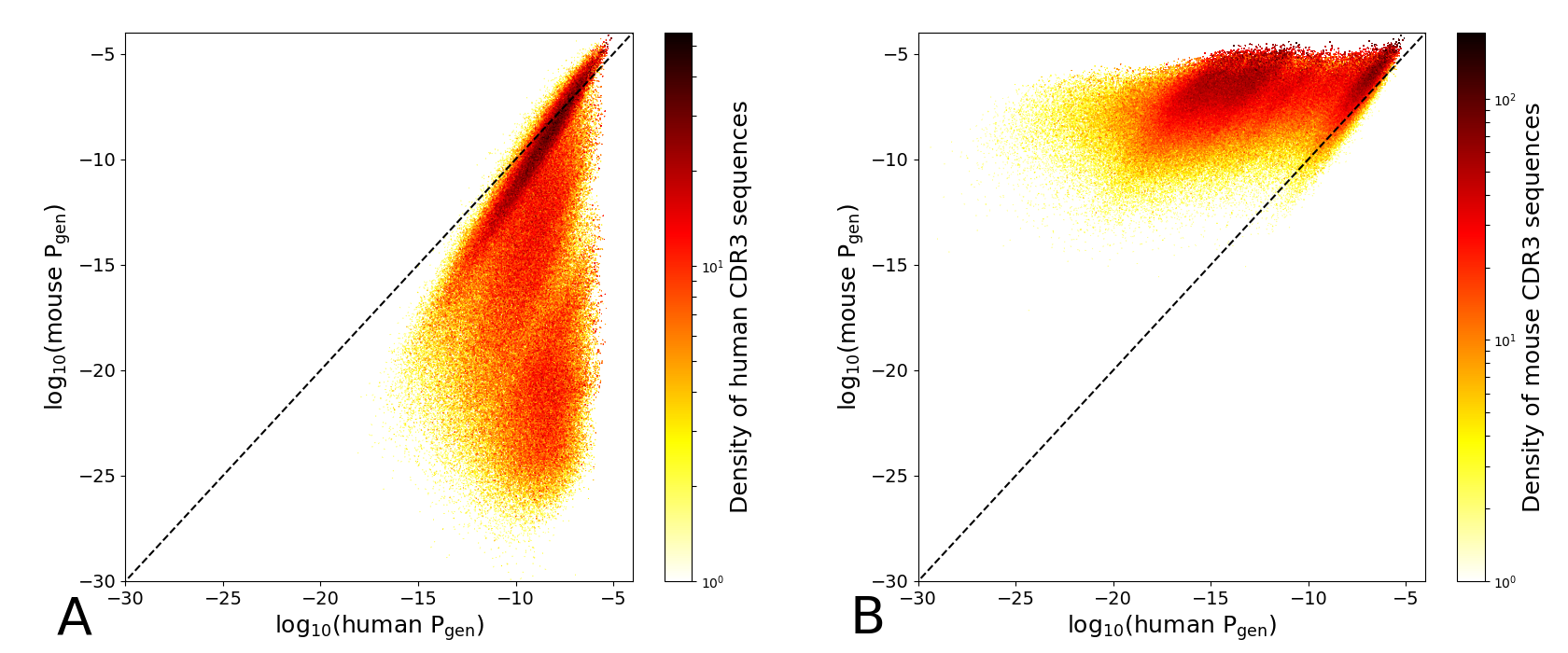}
\caption{\ZS{Scatter plots of CDR3 sequence repertoires across their $P_{\rm gen}$ values as determined by a human TRB model or a mouse TRB model. The sequence repertoires are Monte Carlo samples from A) a human TRB model or B) a mouse TRB model. Projections of the scatter plots onto the two axes reproduce the distributions displayed in Fig.\,\ref{human_mouse_hist}. Models are based on data from \citet{Emerson2017} for human TRB and \citet{Sethna2017} for mouse TRB.}
\label{human_mouse_scat}}
\end{figure}

\YE{The sequences that are produced in one species have substantially lower probability of being generated in the other species (Fig.\,\ref{human_mouse_hist}). The effect is strongest for finding human sequences in a mouse repertoire (compare the black dashed curve with red solid curve): the bulk of the human sequences have extremely low generation probabilities in the mouse model. The effect is less strong for finding mouse sequences in a human repertoire (compare the red dashed curve with the black solid curve):  a small fraction of the mouse sequences have generation probabilities that are as high as the highest generation probabilities of human sequences. The results are not symmetric - while the mouse TRBs processed using the human model have a distinct bi-model distribution, the human TRBs have a very flat and low mouse generation probabilities.} \ZS{Furthermore, mouse TRB sequences always have a non-zero probability of being generated in a human TRB context, however 27.4$\%$ of human TRB sequences have $P_{\rm gen} = 0$ as defined by a mouse TRB model. This asymmetry is primarily due to differences in the insertion profiles (humans may have many more inserted N1 and N2 nucleotides) and by extension CDR3 length.} \YE{Nonetheless, these results suggest that there will be a non-negligible amount of sharing, entirely due to chance statistics, of CDR3 sequences between mouse and human repertoires. A more detailed view of this structure can be seen in a scatter plot of the generation probabilities between the two (Fig.\,\ref{human_mouse_scat}). While there are many sequences with high generation probabilities in both the actual generative model and the cross species model, the cross species generation probabilities are much more variable and span many orders of magnitude, without much correlation to the correct species model.}

\section{Generation probability distributions from additional pathogen response datasets}

\begin{figure}
\begin{center}
\includegraphics[width=.8\linewidth]{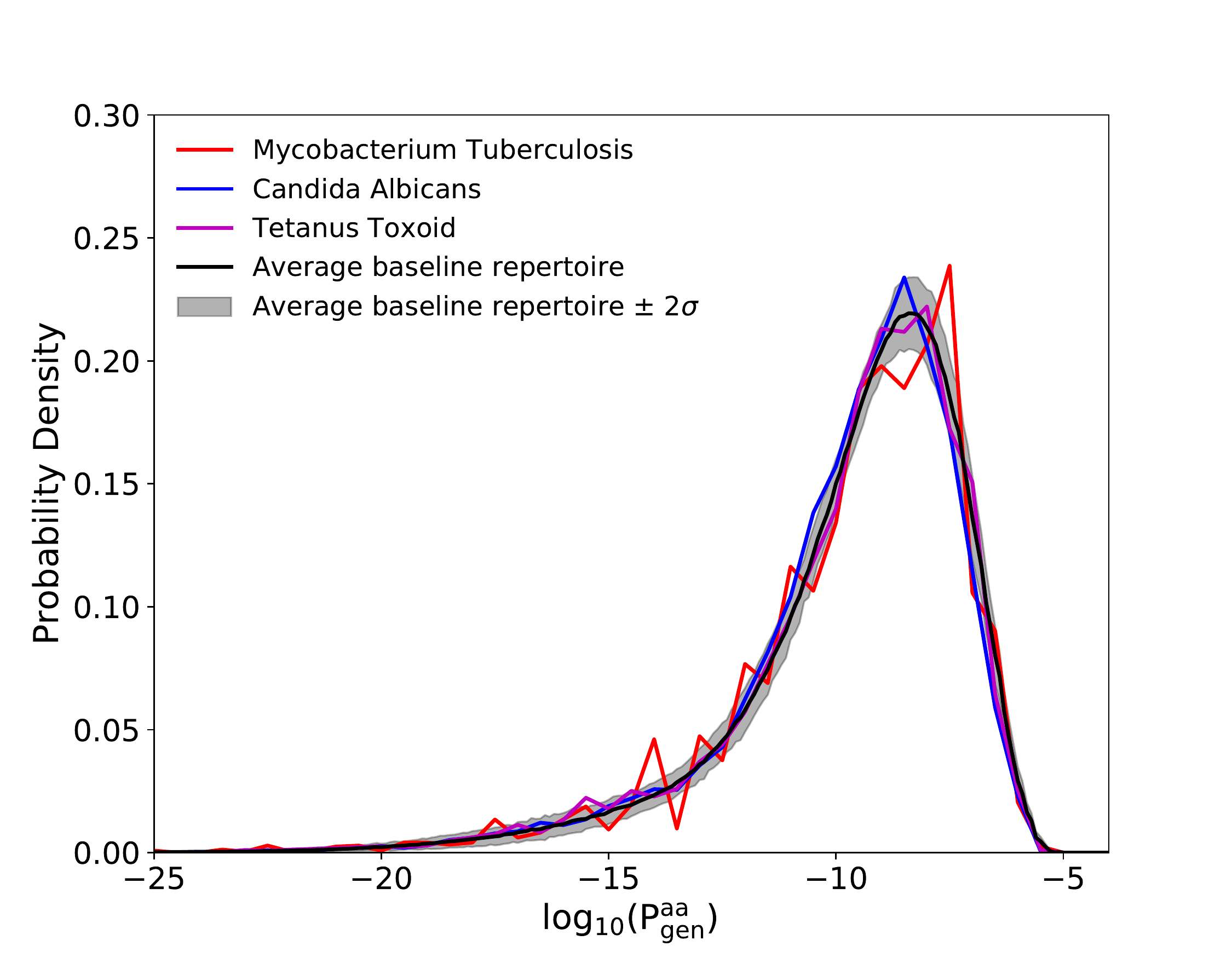}
\caption{\YE{$P_{\rm gen}$ distributions for human CD4+ T cell repertoires that have been incubated with three different pathogens (\citet{Becattini2015}): the fungus Candida Albicans (CA), the bacterium Mycobacterium Tuberculosis (MT), and a toxin protein Tetanus Toxoid (TT). For comparison, the background distribution from human peripheral blood TRB sequences from \citet{Emerson2017} (with its two sigma variation across multiple individuals) is also plotted. The plotted curves are averages over data from individual donors. The sizes of the responsive T cell repertoires are quite variable: the CA dataset has 39934 clonotypes from 5 donors, the TT dataset has 26573 clonotypes from 4 donors, and the MT dataset has 5082 clonotypes from 2 donors. The generation model was infered from \citet{Emerson2017}.}}
\label{Becattini_pgen_dists}
\end{center}
\end{figure}

\YE{In the main text, we displayed the distribution of generative probabilities for T cells known to respond to various pathogens, and even specific epitopes of particular pathogens. The T cell sequences are taken from databases that compile results from multiple experiments. We found that these distributions were, within statistical noise, indistinguishable from the background $P_{\rm gen}$ distribution of PBMCs drawn from the blood. In other words, it would seem that there is no correlation between ease of generation of a T cell and its likelihood to respond to a particular pathogen or epitope. A defect of this analysis is that  the database agglomerates sequences from different experimental protocols, so that there is no way of knowing what biases might have affected the inclusion of any given sequence in the database. Obviously, it would be better to do a single well-controlled experiment in which T cells from a single donor are stimulated to expand by selected pathogens, and the expanded T cells sequenced. Such an experiment was reported by \citet{Becattini2015} several years ago. In their experiment, CD4+ helper T cells were separated from peripheral blood samples, autologous monocytes from the same samples were incubated with three different pathogens (a fungus, a bacterium, and a toxin) in order to load pathogen epitopes, and helper T cell subsamples (typically containing several million T cells, and hundreds of thousands of clonotypes) were incubated with the prepared monocytes (this was done independently for samples from several donors). The T cells in the various samples that had proliferated under this treatment were separated out (typically yielding millions of cells) and their TRB sequences obtained using the Adaptive Biotechnology genomic DNA protocol. The result is a collection of lists of clones (defined by CDR3 amino acid sequence) from the blood of individual donors that can be said to have expanded under stimulation by the three different pathogens. The responses obtained in this way are quite polyclonal, with a few thousand clonotypes in each list of responding clones (the polyclonality perhaps being due to the fact that stimulation is with preparations of whole pathogens, as opposed to particular pathogen peptides). The $P_{\rm gen}$ distributions of the pathogen--responsive clones for different individuals and pathogens are plotted in Fig.\,\ref{Becattini_pgen_dists}. They are indistinguishable from the background $P_{\rm gen}$ distribution derived from blood samples of healthy individuals, which is also plotted (along with its two-sigma variance across a population of individuals) for reference. These data further strengthen the conclusion that pathogen response activity is uncorrelated with $P_{\rm gen}$. }

\end{document}